\documentclass[12pt]{article}
\usepackage[dvips]{graphicx}
\usepackage{cite}

\makeatletter
\newtoks\@stequation

\def\subequations{\refstepcounter{equation}%
  \edef\@savedequation{\the\c@equation}%
  \@stequation=\expandafter{\theequation}
  \edef\@savedtheequation{\the\@stequation}
  \edef\oldtheequation{\theequation}%
  \setcounter{equation}{0}%
  \def\theequation{\oldtheequation\alph{equation}}}

\def\endsubequations{%
  \ifnum\c@equation < 2 \@warning{Only \the\c@equation\space subequation
    used in equation \@savedequation}\fi
  \setcounter{equation}{\@savedequation}%
  \@stequation=\expandafter{\@savedtheequation}%
  \edef\theequation{\the\@stequation}%
  \global\@ignoretrue}

\def\eqnarray{\stepcounter{equation}\let\@currentlabel\theequation
\global\@eqnswtrue\m@th
\global\@eqcnt\z@\tabskip\@centering\let\\\@eqncr
$$\halign to\displaywidth\bgroup\@eqnsel\hskip\@centering
     $\displaystyle\tabskip\z@{##}$&\global\@eqcnt\@ne
      \hfil$\;{##}\;$\hfil
     &\global\@eqcnt\tw@ $\displaystyle\tabskip\z@{##}$\hfil
   \tabskip\@centering&\llap{##}\tabskip\z@\cr}
\makeatother

\def\cerenkov{$\check{\rm C}$erenkov~}

\def\density{{\rm g/cm}^3}
\def\km{\rm km}
\def\deg#1{$#1^{\circ}$}

\def\sis{\sin ^2 2 \theta_{\rm \footnotesize SOL}}
\def\sia{\sin ^2 2 \theta_{\rm \footnotesize ATM}}
\def\sir{\sin ^2 2 \theta_{\rm \footnotesize RCT}}
\def\dmns{\delta_{\rm \footnotesize MNS}}

\def\lsis{$\sin ^2 2 \theta_{\rm \footnotesize SOL}$~}
\def\lsia{$\sin ^2 2 \theta_{\rm \footnotesize ATM}$~}
\def\lsir{$\sin ^2 2 \theta_{\rm \footnotesize RCT}$~}
\def\ldmns{$\delta_{\rm \footnotesize MNS}$~}

\def\PR#1#2#3{Phys. Rev. {\bf #1}, #2 (#3)}
\def\PRL#1#2#3{Phys. Rev. Lett. {\bf #1}, #2 (#3)}
\def\PL#1#2#3{Phys. Lett. {\bf #1}, #2 (#3)}
\def\NL#1#2#3{Nucl. Phys. {\bf #1}, #2 (#3)}

\def\PTP#1#2#3{Prog. Theor. Phys. {\bf #1}, #2 (#3)}

\def\PRD#1#2#3{Phys. Rev. {\bf D#1},~#2 (#3)}

\def\EPJC#1#2#3{Eur. Phys. J. {\bf C#1} #2 (#3)}
\def\PLB#1#2#3{Phys. Lett. {\bf B#1} #2 (#3)}

\def\eqref#1{eq.~(\ref{#1})}
\def\figref#1{Fig.~\ref{#1}}

\def\etal{{\it et al.}}
\def\ibid{{\it ibid.~}}

\def\simgt{\lower.5ex\hbox{$\; \buildrel > \over \sim \;$}}
\def\simlt{\lower.5ex\hbox{$\; \buildrel < \over \sim \;$}}

\title{Physics potential of T2KK:\\
An extension of the T2K neutrino oscillation experiment with a far detector in Korea
}

\author{Kaoru Hagiwara$^{1}$, 
Naotoshi Okamura$^2$\thanks{e-mail:~okamura@yukawa.kyoto-u.ac.jp} , 
and 
Ken-ichi Senda$^{1}$\thanks{e-mail:~senda@post.kek.jp}
\\ \\
{\it \small $^1$KEK Theory Division, and the Graduate University
for Advanced Studies (SOKENDAI),}\\
{\it \small Tsukuba, 305-0801 Japan} \\ 
{\it \small $^2$Yukawa Institute for Theoretical Physics, Kyoto University, }\\
{\it \small Kyoto, 606-8502 Japan}
}
\date{}
\begin{document}
\maketitle
\vspace{-9.5cm}
\begin{flushright}
YITP-06-29 
\hspace*{3ex}
KEK-TH-1091
\hspace*{3ex}
hep-ph/0607255
\end{flushright}
\vspace{ 9.5cm}
\vspace{-2.0cm}

\begin{abstract}
We study physics potential of placing a far detector in the east coast
of Korea,
where the off-axis neutrino beam from J-PARC at Tokai village for the T2K
project has significant intensity at a few GeV range.
In particular, we examine the capability of determining the mass hierarchy
pattern and the CP phase of the lepton-flavor-mixing matrix
when a 100 kt water \cerenkov detector is placed at various locations in Korea
for the off-axis beam (OAB) of \deg{2.5} and \deg{3.0} at the
Super-Kamiokande site. 
The best results are found for a combination of \deg{3.0} OAB at SK
($L = 295 \km$) and \deg{0.5} OAB at $L = 1000 \km$,
where the mass hierarchy pattern can be determined at 3-$\sigma$ level
for $\sir  \simgt 0.05$ $(0.06)$ when the hierarchy is normal
(inverted), after 5 years of running ($5 \times 10^{21}$ POT).
We also find that the leptonic CP phase, \ldmns, can be constrained uniquely,
without invoking anti-neutrino beams,
as long as the mass hierarchy pattern is determined.
Those results are obtained by assuming that the charged current
quasi-elastic events can be separated from the other backgrounds with high
efficiency, the neutrino energy can be reconstructed with a hundred
MeV uncertainty, and the earth matter density along the baseline can be
determined with $3\%$ accuracy.
\end{abstract}

\newpage
\section{Introduction}
All the experimental results of the neutrino flavor oscillation 
\cite{Pon}
are consistent with 3 neutrinos
except for the LSND \cite{LSND} experiment. 
Among the 9 parameters of the three neutrino model, 6 parameters can
be measured by neutrino oscillation experiments:
2 mass-squared differences, 3 mixing angles and 1 CP violating phase.
As of 2006 spring,
we have already known 2 mass-squared differences and 2 mixing angles
from the atmospheric neutrino experiments \cite{atomos-result}, the
long-base-line (LBL) neutrino oscillation experiments \cite{k2k,
minos}, the solar neutrino experiments \cite{sol-result1,sol-result2},
and the reactor neutrino experiments \cite{kamland, chooz}.
However, the sign of the larger mass squared-difference ($m^2_3 - m^2_1$),
one of the 3 mixing angles ($\theta_{\rm \footnotesize RCT}$),
and the CP phase $(\dmns)$ have not been measured yet.
The tasks of the future neutrino oscillation experiments are not only
to confirm the 3 neutrino model, but also to measure those unknown
parameters of the model.
\par
Here we focus our attention on 
one of the future neutrino experiments, the Tokai-to-Kamioka
experiment (T2K) \cite{T2K}
which will start in 2008.
The center of the T2K neutrino beam from J-PARC \cite{j-parc} at Tokai 
village will go through underground beneath Super-Kamiokande (SK), and 
reach the sea level east of Korean shore.
At the baseline length $L = 295$ km away from Tokai village
the upper side of the beam at $2^\circ$ to $ 3^\circ$ off-axis angle is observed at SK,
and the lower side of the same beam at $0.5^\circ$ to $3.0^\circ$ off-axis angle
can be observed in Korea
\cite{hagiwara-see-saw}.
An additional far detector in Korea can probe the neutrino oscillation
at a baseline length ($L$) of 1000 to 1200 km away from the Tokai village
\cite{hagiwara-see-saw, t2kk, t2kr-l}.
The most welcome feature of this two detector system has been identified \cite{t2kr-l}
as the relative hardness of the neutrino beam energy spectrum at a smaller off-axis angle,
which can be observed in the east coast of Korea at $L \sim 1000$ km.
Accordingly, it is possible to arrange such that a far detector in Korea probe
the oscillation at around the same oscillation phase, $\delta m^2 L / 2E$, at Kamioka.
The difference between the resulting two measurements should then come from the difference in the
matter effects, which can be a factor of three larger in Korea.
In Ref.~\cite{t2kr-l}, we showed that the two detector system can resolve the mass hierarchy
ambiguity by making use of the strong matter effects on the $\nu_{\mu} \to \nu_e$ transition probability
\cite{narayan99, barger01, t2b, supernova} if a 100 kt level water \cerenkov detector is
placed in the east coast of Korea and if the third mixing angle is not too small.
It has further been shown in \cite{t2kr-l} that the leptonic CP phase, \ldmns, can be
uniquely constrained once the hierarchy pattern is determined.
\par
In this paper we present details of our findings in Ref.~\cite{t2kr-l}.
The paper is organized as follows.
In section 2, we review the formalism of the neutrino oscillation
including the matter effect, as well as the constraints in the model parameters from the
present experimental results.
In section 3, we briefly introduce the T2K experiment and discuss
merits of the Tokai-to-Kamioka-and-Korea (T2KK) proposal,
where an additional large detector is placed in Korea along the T2K neutrino beam baseline.
In section 4, we explain details of our analysis method where we select
the charged current quasi-elastic events, and a $\chi^2$ function is proposed
that takes into account statistical and systematic uncertainties in the experiment,
as well as a part of possible background contaminations.
In section 5, we show the results of our numerical calculation
on the determination of the neutrino mass hierarchy.
In section 6, we present our studies on the determination of the CP phase.
In section 7, we summarize our findings and give some discussions on
the possibility of expanding the T2KK two detector system,
and the necessity of further studies on backgrounds which we could not include in the present analysis.
\section{{\large Oscillation formulae under the experimental constraints}}
\par
The neutrino flavor eigenstates, $\left| \nu_{\alpha} \right\rangle$ ($\alpha = e, \mu, \tau $ ),
are related to the mass eigenstates $\left| \nu_{i} \right\rangle$ ($i = 1, 2, 3$)
through the lepton-flavor-mixing matrix, or the Maki-Nakagawa-Sakata (MNS) matrix \cite{mns}
\begin{equation}
\left| \nu_{\alpha} \right\rangle =  U_{\alpha i} \left|\nu{_i} \right\rangle \,.
\label{eq:state}
\end{equation}
The probability that an initial flavor eigenstate $\nu_{\alpha}$ with energy $E$ is observed as
a flavor eigenstate $\nu_{\beta}$ after traveling a distance $L$ in the vacuum is expressed as
\begin{equation}
P_{\nu_{\alpha} \to \nu_{\beta}} =
| U_{\beta 1}U_{\alpha1}^*  + U_{\beta 2} U_{\alpha2}^* e^{-i\Delta_{12}}
+ U_{\beta 3} U_{\alpha3}^* e^{-i\Delta_{13}}|^2  \,,
\label{eq:prob}
\end{equation}
where the phase $\Delta_{ij}$ is
\begin{equation}
\Delta_{ij} = \frac{m^2_j - m^2_i}{2E} L \simeq 2.534
\frac{ (m^2_j - m^2_i) [{\rm eV^2}]}{E {\rm [GeV]}} L [\km ] \,.
\label{eq:Deltaij}
\end{equation}
Eq.~(\ref{eq:prob}) tells us
that neutrino oscillation experiments measure the
2 mass-squared differences and
the lepton-number conserving contributions of the
MNS matrix elements,
which can be parametrized by
3 mixing angles and 1 CP violating phase.
\par
The present neutrino oscillation experiments are each sensitive to only
one of the 2 mass-squared differences, and the amplitude of each oscillation
probability can be expressed in terms of the MNS matrix elements.
The atmospheric neutrino oscillation experiments \cite{atomos-result}
and the long-base line (LBL) neutrino oscillation experiments,
K2K\cite{k2k} and MINOS \cite{minos}, which measure the $\nu_{\mu}$ survival probability,
are sensitive to the magnitude of the larger mass-squared
difference.
The constraints on the mass-squared difference and the amplitude are \cite{atomos-result, k2k, minos}
\begin{subequations}
\label{eq:atm-data}
\begin{eqnarray}
&1.5\times10^{-3} {\mbox{{eV}}}^2 < |m_3^2 - m_1^2| < 3.4 \times10^{-3} {\mbox{{eV}}}^2 \,,\\
&\sia > 0.92  \label {eq:sia-data}\,,
\end{eqnarray}
\end{subequations}
each at the 90$\%$ confidence level.
The solar neutrino oscillation experiments \cite{sol-result1,sol-result2} and the KamLAND \cite{kamland} experiment,
which measure the $\nu_e$ and $\bar{\nu}_e$ survival probability, respectively,
are sensitive to the smaller mass-squared difference.
The present constraints can be expressed as \cite{sol-result2}
\begin{subequations}
\label{eq:sol-data}
\begin{eqnarray}
&m^2_2 - m^2_1 = \left(8.0 \pm 0.3 \right)\times 10^{-5} {\mbox{{eV}}}^2 \,,\\
&\sin^2\theta_{\rm \footnotesize SOL} = 0.30 \pm 0.03\,. 
\label{eq:sol-angle}
\end{eqnarray}
\end{subequations}
In the solar neutrino experiments, the sign of $m^2_2 - m^2_1$ is determined by
the matter effect in the sun \cite{matter-effect, msw}.
The reactor $\bar{\nu}_e$ experiments at $L \sim 1$ km are
sensitive to the oscillation with the larger mass-squared difference.
No reduction of the $\bar{\nu}_e$ survival probability has been observed,
and the CHOOZ experiment gives the upper limit on the amplitude \cite{chooz};
\begin{subequations}
\label{eq:rct-data}
\begin{eqnarray}
&\sir < 0.20 \mbox{{~  for  ~}}
| m_3^2 - m_1^2 | = 2.0\times10^{-3}\mbox{{eV}}^2\,,\\
\label{eq:rct-just}
&\sir < 0.16 \mbox{{~  for  ~}}
| m_3^2 - m_1^2 | = 2.5\times10^{-3}\mbox{{eV}}^2\,,
\\
&\sir < 0.14 \mbox{{~  for  ~}}
| m_3^2 - m_1^2|
= 3.0\times10^{-3}\mbox{{eV}}^2\,,
\end{eqnarray}
\end{subequations}
at the 90$\%$ confidence level.
\par
These observed amplitudes can be identified with
the MNS matrix elements as follows \cite{hagiwara-okamura}.
In the atmospheric neutrino experiments and the CHOOZ reactor experiment,
$|\Delta_{13}| \sim 1 \gg \Delta_{12}$ is satisfied. Therefore we can set $e^{-i \Delta_{12}} \sim 1$
in \eqref{eq:prob} and we find the following relations;
\begin{eqnarray}
|U_{\mu3}|^2 &=& \sin^2 \theta_{\rm \footnotesize ATM} \nonumber \,, \\
|U_{e3}|^2 &=& \sin^2 \theta_{\rm \footnotesize RCT} \,. 
\label{eq:mix1}
\end{eqnarray}
As for the solar neutrino experiments and the KamLAND experiment,
where the $\Delta_{12} \sim 1$ region is probed,
the terms with $\Delta_{13}$ oscillate rapidly within the experimental resolution of $L/E$.
After averaging out the $\Delta_{13}$ contribution, and by neglecting the
term of order $|U_{e3}|^2$, which is constrained to be smaller than about 0.04 by the
CHOOZ experiment \cite{chooz}, \eqref{eq:rct-just},
we obtain the relation;
\begin{eqnarray}
4|U_{e1} U_{e2}|^2 = \sin^2 2\theta_{\rm \footnotesize SOL} \,.
\label{eq:mix2}
\end{eqnarray}
These simple identifications, eqs.~(\ref{eq:mix1}) and (\ref{eq:mix2}),
are found to give a reasonably good description of the present data
in dedicated studies \cite{lisi}
of the experimental constraints in the three neutrino model.
In this paper we parametrize the MNS matrix in terms of the three positive numbers,
$\sin^2 \theta_{\rm \footnotesize ATM}$, $\sin^2 \theta_{\rm \footnotesize RCT}$,
and \lsis with the identification of eqs.~(\ref{eq:mix1}) and (\ref{eq:mix2}), 
respectively,
and the CP phase
\begin{equation}
\dmns = - \arg U_{e3} \,.
\end{equation}
This convention \cite{hagiwara-okamura} allows us to express the MNS matrix
in terms of the three observed amplitudes, directly.
\par
The neutrino oscillation through the earth is complicated by the fact that
$\nu_e$ and $\bar{\nu}_e$ have the extra $W$-boson exchange interactions
with electrons in the matter \cite{matter-effect}.
The effect is small at low energies, and an approximation of keeping only
the first and second order corrections
in the matter effect and the smaller mass-squared difference has been
found useful for analyzing the LBL experiments at sub
GeV to a few GeV range \cite{arafune97, koike05}.
Although we evaluate the oscillation probabilities numerically
in this report,
the following analytic expressions are found to be useful
for the T2KK setup where the earth
matter effects can be treated pertubatively.
For the $\nu_{\mu}$ survival probability,
an accurate analytic expressions is found
by retaining only those terms linear in
$\Delta_{12}$ and the matter effect term
\cite{koike05, t2kr-l}:
\begin{equation}
P_{\nu_{\mu} \rightarrow \nu_{\mu}} 
=  1 - \sin^2 2\theta_{\rm \footnotesize ATM}\left( 1 + A^{\mu} \right)
       \sin^2 \left( \displaystyle\frac{\Delta_{13}}{2} + B^{\mu} \right) \,.
\label{eq:p-numu-numu}
\end{equation}
Here $A^{\mu}$ and $B^{\mu}$ 
are the corrections to the amplitude and the oscillation phase, respectively,
which are linear in the smaller mass squared difference and the matter effect term.
These can be expressed as,
\begin{subequations}
\label{eq:AB-mu}
\begin{eqnarray}
A^{\mu} &=& - \displaystyle\frac{aL}{\Delta_{13}E} 
\displaystyle\frac{ 1 -  2 \sin^2 \theta_{\rm \footnotesize ATM}}
{\cos^2 \theta_{\rm \footnotesize ATM}}
\sin^2 \theta_{\rm \footnotesize RCT} \nonumber \\
& \sim & - 0.005
\left( 1 -  2 \sin^2 \theta_{\rm \footnotesize ATM} \right)
\displaystyle\frac{\pi}{\Delta_{13}}
\displaystyle\frac{L}{295{\rm km}}
\left ( \frac{\sir}{0.10} \right ) \,,
\label{eq:A-mu}\\
B^{\mu} &=& \displaystyle\frac{aL}{4E} 
\displaystyle\frac{1 -  2 \sin^2 \theta_{\rm \footnotesize ATM}}
{\cos^2 \theta_{\rm \footnotesize ATM}}
\sin^2 \theta_{\rm \footnotesize RCT}
\nonumber \\
  && \hspace{-1.0cm} - \displaystyle\frac{\Delta_{12}}{2}
\left(  \cos^2 \theta_{\rm \footnotesize SOL} 
       + \tan^2 \theta_{\rm \footnotesize ATM}\sin^2 \theta_{\rm \footnotesize SOL}\sin^2\theta_{\rm \footnotesize RCT}
       - \tan \theta_{\rm \footnotesize ATM}\sin 2\theta_{\rm \footnotesize SOL}\sin \theta_{\rm \footnotesize RCT}\cos\dmns
\right) \nonumber \\
&\sim& -
\left[0.037  - 0.008 
\left(
\displaystyle\frac{\sin^2 2\theta_{\rm \footnotesize RCT}}{0.10}
\right)^{1/2}
\cos \dmns 
\right]
\displaystyle\frac{|\Delta_{13}|}{\pi} \,.
\label{eq:B-mu}
\end{eqnarray}
\end{subequations}
Here the term $a$ gives the contribution of the extra potential for $\nu_e$
\begin{equation}
a = 2\sqrt{2} G_F E n_e \approx 7.56 \times 10^{-5} {\rm eV}^2
\left( \frac{\rho}{{\rm g/cm}^3}\right)
\left( \frac{E}{{\rm GeV}} \right) \,,
\label{eq:matter-effect}
\end{equation}
where $n_e$ is the electron
number density and $\rho$ is the
matter density.
In the second lines of
eqs.~(\ref{eq:A-mu}), (\ref{eq:B-mu}),
we employ the mean values of the atmospheric and
the solar oscillation parameters in 
eqs.~(\ref{eq:atm-data}) and (\ref{eq:sol-data}),
and $\rho = 3.0 \density$.
The simple analytic expressions of eqs.~(\ref{eq:p-numu-numu}) and (\ref{eq:AB-mu})
reproduces the survival probability accurately with 1$\%$ error
throughout the parameter range explored in this report,
except where the probability vanishes.
\par
According to eqs.~(\ref{eq:A-mu}), (\ref{eq:sia-data}), and (\ref{eq:rct-just}),
the magnitude of $A^{\mu}$
around $|\Delta_{13}| = \pi$ should be smaller than
about $2 \times 10^{-3}$ ($ 8 \times 10^{-3}$) at $L = 295$ km (1000 km),
and hence
the amplitude of the $\nu_{\mu}$ survival probability is not affected much by
the matter effect.
Therefore, we can measure \lsia rather uniquely from the $\nu_\mu$
disappearance probability independent of the neutrino mass hierarchy
and the other unconstrained parameters.
The phase-shift term $B^{\mu}$ affects the measurement of $|m^2_3 - m^2_1|$,
whose preferred value grows (decreases) for the normal (inverted) hierarchy,
and the magnitude of the shift can be about to $2 \%$ ($3 \%$) when
$\cos \dmns = 1$  ($-1$) for $\sir = 0.1$.
In other words, unless we determine the hierarchy,
a few percent level of uncertainty should remain as a systematic error
of $|m^2_3 - m^2_1|$.
\par
For the $\nu_{\mu} \to \nu_e$ transition probability,
we need not only the linear terms of $\Delta_{12}$ and $a$ but also
their quadratic terms to obtain a good approximation.
We find
\begin{equation}
P_{\nu_{\mu} \rightarrow \nu_e} = 
4 \sin^2 \theta_{\rm \footnotesize ATM} \sin^2 \theta_{\rm \footnotesize RCT}
\left\{
 \left( 1 + A^{e} \right)
  \sin^2 \left( \displaystyle\frac{\Delta_{13}}{2} \right)
+B^e \sin \Delta_{13}
\right\}
+C^e
 \,,
\label{eq:p-numu-nue}
\end{equation}
where 
$A^e$, $B^e$, and $C^e$ are the correction terms:
\begin{subequations}
\label{eq:AB-e}
\begin{eqnarray}
A^e &=& \displaystyle\frac{aL}{\Delta_{13}E} 
        \cos 2\theta_{\rm \footnotesize RCT} 
        -\displaystyle\frac{\Delta_{12}}{2}
	\displaystyle\frac{\sin2\theta_{\rm \footnotesize SOL}}{\tan
	\theta_{\rm \footnotesize ATM} \sin\theta_{\rm \footnotesize
	RCT}}\sin \dmns 
	\left( 1 + \displaystyle\frac{aL}{2\Delta_{13}E}\right)
	\nonumber\\
	&&+ \displaystyle\frac{\Delta_{12}}{4}
	    \left( \Delta_{12} + \displaystyle\frac{aL}{2E} \right)
	    \left(
	    \displaystyle\frac{\sin2\theta_{\rm \footnotesize SOL}}{\tan
	      \theta_{\rm \footnotesize ATM} \sin\theta_{\rm \footnotesize
	     RCT}}\cos \dmns 
	     - 2 \sin^2 \theta_{\rm \footnotesize SOL}
	     \right) \nonumber \\
	 &&-\displaystyle\frac{1}{2}\left( \displaystyle\frac{aL}{2E} \right)^2
	     +\frac{3}{4}
	     \left( \displaystyle\frac{aL}{\Delta_{13}E} \right)^2
\label{eq:A-e}
\\
 &\sim& 0.11
\displaystyle\frac{\pi}{\Delta_{13}}
\displaystyle\frac{L}{295{\rm km}}
\left( 1 - \displaystyle\frac{\sir}{2} \right)
-0.014
\left(\displaystyle\frac{L}{295{\rm km}}\right)^2
+
0.0087
\left(\displaystyle\frac{\pi}{\Delta_{13}}
\displaystyle\frac{L}{295{\rm km}}\right)^2
\nonumber \\
&&
-
0.29
\left( \displaystyle\frac{0.10}{\sin^2 2\theta_{\rm \footnotesize RCT}} \right)^{1/2}
\sin \dmns
\left(
\displaystyle\frac{|\Delta_{13}|}{\pi}
\pm 0.054 \displaystyle\frac{L}{295{\rm km}}
\right) 
\nonumber \\
&&
+0.015 \left[
\left( \displaystyle\frac{0.10}{\sin^2 2\theta_{\rm \footnotesize
 RCT}} \right)^{1/2}
\cos \dmns
- 0.11
\right]
\left(
\displaystyle\frac{|\Delta_{13}|}{\pi}
+
1.7
\displaystyle\frac{L}{295{\rm km}}
\right)
\displaystyle\frac{|\Delta_{13}|}{\pi}
\,,
\label{eq:A-e+}
\\
B^e &=& - \displaystyle\frac{aL}{4E}\cos 2\theta_{\rm \footnotesize RCT}
         + \displaystyle\frac{\Delta_{12}}{4}
                       \left(\displaystyle\frac{\sin2\theta_{\rm
			\footnotesize SOL}}{\tan \theta_{\rm
			\footnotesize ATM} \sin\theta_{\rm
			\footnotesize RCT}}\cos \dmns - 2 \sin^2
			\theta_{\rm \footnotesize SOL} \right)
		    \left( 1 + \displaystyle\frac{aL}{2\Delta_{13}E}\right) 
\nonumber \\
&&+ \displaystyle\frac{\Delta_{12}}{8}
	    \left(
	      \Delta_{12} + \displaystyle\frac{aL}{2E}
	    \right)
	    \displaystyle\frac{\sin2\theta_{\rm \footnotesize SOL}}{\tan
	      \theta_{\rm \footnotesize ATM} \sin\theta_{\rm \footnotesize
	     RCT}}\sin \dmns 
    - \displaystyle\frac{1}{\Delta_{13}}
    \left(  \displaystyle\frac{aL}{2E}    \right)^2
    \label{eq:B-e}
\\
 &\sim&
- 0.080
\left( \displaystyle\frac{L}{295{\rm km}} \right)
\left(1 - \displaystyle\frac{\sir}{2} \right)
-
0.0091\displaystyle\frac{\pi}{\Delta_{13}}
\left( \displaystyle\frac{L}{295{\rm km}} \right)^2
\nonumber \\
&&
+0.15
\left[
\left( \displaystyle\frac{0.10}{\sir} \right)^{1/2} \cos \dmns
- 0.11
\right] 
\left(
\displaystyle\frac{|\Delta_{13}|}{\pi}
\pm 0.054\displaystyle\frac{L}{295{\rm km}}
 \right)
\nonumber \\
&&
+0.0072
\left( \displaystyle\frac{0.10}{\sin^2 2\theta_{\rm \footnotesize
 RCT}} \right)^{1/2}
\sin \dmns
\left(
\displaystyle\frac{|\Delta_{13}|}{\pi}
+
1.7
\displaystyle\frac{L}{295{\rm km}}
\right)
\displaystyle\frac{|\Delta_{13}|}{\pi}
\label{eq:B-e+}\,, \\
C^e &=&
\displaystyle\frac{\Delta_{12}^2}{4}
\sin^2 2\theta_{\rm \footnotesize SOL}
\cos^2 \theta_{\rm \footnotesize ATM}
-
\displaystyle\frac{\Delta_{12}}{2}
\displaystyle\frac{aL}{2E}
\sin 2\theta_{\rm \footnotesize SOL}
\sin 2\theta_{\rm \footnotesize ATM}
\sin \theta_{\rm \footnotesize RCT}
\cos \dmns
\label{eq:C-e}
\\
&&
+\left(
\displaystyle\frac{aL}{2E}
\right)^2
\sin^2 \theta_{\rm \footnotesize RCT}
\sin^2 \theta_{\rm \footnotesize ATM}
\nonumber \\
&\sim&
0.0011
\left(\displaystyle\frac{\Delta_{13}}{\pi} \right)^2
-
0.0013
\displaystyle\frac{|\Delta_{13}|}{\pi}
\displaystyle\frac{L}{295 {\rm km}}
\left(
\displaystyle\frac{\sir}{0.10}
\right)^{1/2}
\cos \dmns
\nonumber
\\
&&
+
0.00036
\left(
\displaystyle\frac{L}{295{\rm km}}
\right)^2
\displaystyle\frac{\sir}{0.10}
\,.
\label{eq:C-term}
\end{eqnarray}
\end{subequations}
Here $\pm$ in the first lines of eqs.~(\ref{eq:A-e+}) and (\ref{eq:B-e+})
follows the sign of $\Delta_{13}$.
The first and second terms
in eqs.~(\ref{eq:A-e}) and (\ref{eq:B-e})
are linear, 
and the other terms including
\eqref{eq:C-e}
are quadratic in $\Delta_{12}$ and $a$ corrections.
The quadratic terms can dominate the probability when
\lsir, which gives the leading term,
is very small.
We find that the above analytic expressions,
eqs.~(\ref{eq:p-numu-nue}) and (\ref{eq:AB-e}),
are useful throughout the
parameter range of this report,
down to $\sir = 0$,
except near the oscillation minimum
where the approximation can give
negative probability.
The largest deviation is found for
$\sir \approx 0.005$,
$\delta_{\rm MNS} \approx 0^{\circ}$
and the inverted hierarchy,
where the analytic formula underestimates
the probability by about $5\%$ for $L = 1000$ km.
\par
The term proportional to $\sin^2 (\frac{\Delta_{13}}{2})$ in \eqref{eq:p-numu-nue}
is the main oscillation term in our approximation,
and $A^e$ shifts the magnitude of the amplitude.
From \eqref{eq:A-e+}, we find that the amplitude of the $\nu_{\mu} \to \nu_e$ 
transition probability
is sensitive to the mass hierarchy pattern,
because the first term changes sign.
Its magnitude increases (decreases) by about $10 \%$ for the normal
(inverted) hierarchy at $L = 295$ km for $\sin \dmns = 0$.
The difference between the two hierarchy cases
grows with $L$ when $L/E$ is fixed at $|\Delta_{13}| \sim \pi$,
reaching about $\pm40\%$ at $L = 1000$ km.
The shift is also sensitive to $\sin \dmns$, which can decrease (increase)
the amplitude by as much as $30 \%$ for $\sin \dmns = 1 (-1)$
when $\sir = 0.1$, independent of the mass hierarchy.
\par
The term proportional to $B^e$ in \eqref{eq:p-numu-nue}
can be re-organized as 
\begin{equation}
P_{\nu_{\mu} \rightarrow \nu_e} \approx 
4 \sin^2 \theta_{\rm \footnotesize ATM} \sin^2 \theta_{\rm \footnotesize RCT}
 \left( 1 + A^{e} \right)
  \sin^2 \left( \displaystyle\frac{\Delta_{13}}{2} +B^e \right)
 +C^e 
 \,,
 \label{eq:factorize}
\end{equation}
just like the $\nu_{\mu}$ survival formula, \eqref{eq:p-numu-numu},
when $|A^e|$ and $|B^e|$are small.
Since the term $B^e$ shifts the oscillation phase in this limit,
we call it as the phase-shift term.
It shifts the oscillation peak energy higher
(lower)
for the normal (inverted) hierarchy,
by about $5 \%$ at $L = 295$ km and by about $20 \%$ at $L = 1000$ km,
for $\cos \dmns = 0$.
The $\cos \dmns$ dependence of the shift is also significant,
which can be $\pm 10\%$ for $\cos \dmns = \pm 1$ if $\sir = 0.1$.
Note, however, the factorized expression \eqref{eq:factorize}
seizes to be a good approximation when $|A^e|$ or $|B^e|$
gets larger than about 0.5,
which happens when $\sir \sim 0.01$;
see eqs.~(\ref{eq:A-e+}) and (\ref{eq:B-e+}).
\par
The above observation inspires a two detector system,
where both detectors can measure $\nu_{\mu} \to \nu_e$ oscillations around
the oscillation maximum ($|\Delta_{13}| \sim \pi$)
but at significantly different base-line lengths. If the magnitude of
the product of
$\sin^2 \theta_{\rm \footnotesize ATM} \sin^2 \theta_{\rm \footnotesize RCT}$
is large enough that the $\nu_{\mu} \to \nu_e$ oscillation is observed
at both detectors,
then the difference of the observed oscillation probabilities at the
two locations determines the
mass hierarchy uniquely, and hence also the
product $\sin^2 \theta_{\rm \footnotesize ATM} \sin^2 \theta_{\rm
\footnotesize RCT}$ rather independent of $\dmns$.
Once the hierarchy is determined, the measurements of the magnitude
and the phase of the $\nu_{\mu} \to \nu_e$ oscillation
measure $\sin \dmns$ and $\cos \dmns$, respectively,
and hence the CP violating phase \ldmns can be determined uniquely.
In the following section, we will find that the T2KK two detector system can indeed satisfy
the above conditions, and our $\chi^2$ analysis based on the exact numerical evaluation
of the oscillation probability supports the simple picture presented in this
section based on the approximate formula eqs.~(\ref{eq:p-numu-numu}) - (\ref{eq:AB-e}).
\section{{\large Merits of detecting the T2K off-axis beam in Korea}}
\begin{figure}
\begin{center}
\includegraphics[angle = 0 , width = 10cm]{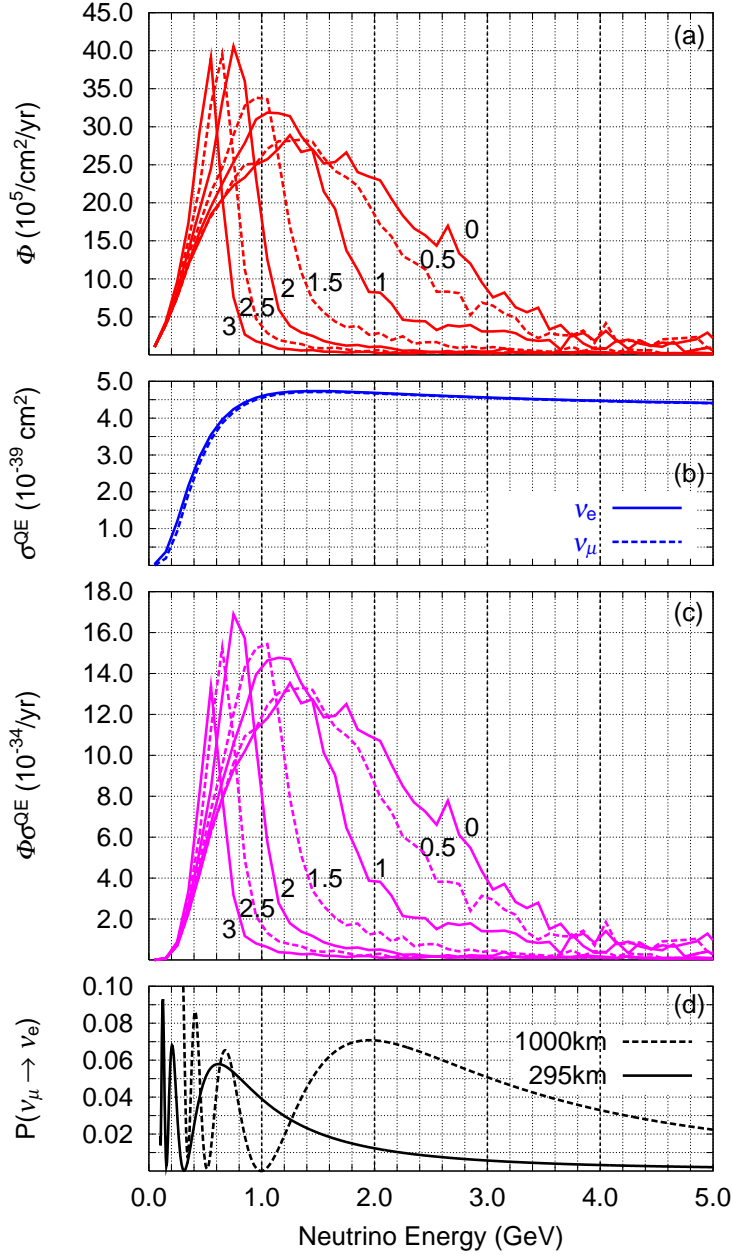}
\end{center} 
\caption{
(a).~The flux of the T2K $\nu_{\mu}$ beam for $10^{21}$ POT/yr
at $L$ = 295 km for various off-axis angles between \deg{0} and \deg{3}. 
(b).~The cross section per nucleon
of the Charged Current Quasi Elastic (CCQE) events for $\nu_e$ and $\nu_{\mu}$
off the water target. 
(c).~The flux at 295 km times $\nu_e$ CCQE cross section.
(d).~ Probability  of $\nu_{\mu} \to \nu_e$ transition at 295 km (solid line)
and that at 1000 km (dashed line) calculated for $m^2_3 - m^2_1 = 2.5
 \times 10^{-3} {\rm eV}^2$,
$m^2_2 - m^2_1 = 8.2 \times 10^{-5} {\rm eV}^2$, $\sia = 1.0$, $\sis =
 0.83$, $\sir = 0.10$,
$\dmns = 0^{\circ}$, and $\rho = 2.8 \density $ for $ L = 295~\km$,
and $\rho = 3.0 \density$ for $L = 1000~\km $.
}
\label{fig:flux}
\end{figure}

Many neutrino oscillation experiments \cite{double-chooz, kaska, daya-bay, braidwood, reno, angra, nova}
are planned to measure \lsir.
Tokai-to-Kamioka (T2K) neutrino-oscillation experiment which will start in 2008 is one of them.
In T2K,  high intensity $\nu_{\mu}$ beam is produced by the proton accelerator which is under construction at
J-PARC in Tokai-village.
These $\nu_{\mu}$'s are shot to Kamioka, 295 km west from Tokai.
Super-Kamiokande (SK) will measure both the $\nu_{\mu} \to \nu_{\mu}$
survival rate and the $\nu_{\mu} \to \nu_e$ transition rate.
The measurement of the $\nu_{\mu} \to \nu_e$ transition  probability is the main purpose of the T2K experiment,
because it tells us the magnitude of $\sin^2 \theta_{\rm \footnotesize RCT}$, see \eqref{eq:p-numu-nue},
the last unmeasured mixing angle of the $3 \times 3$ MNS matrix.
In order to observe the signal of $\nu_{\mu} \to \nu_e$ clearly,
the neutrino beam should satisfy the following conditions.
\begin{enumerate}
  \item The neutrino energy near the oscillation maximum ($|\Delta_{13}| \sim \pi$) at $L = 295$km is
  expected to be around 0.5 GeV to 0.7 GeV
  according to the present experimental bound, \eqref{eq:atm-data}.
The $\nu_{\mu}$ flux should hence be large in this energy region.
 \item High energy neutrinos produce $\pi^0$'s via neutral current,
which become background to the $\nu_e$ Charged Current Quasi-Elastic (CCQE) events.
Therefore, the flux of $\nu_{\mu}$ beam should be small at  high energies.
\end{enumerate}
T2K adopts the off-axis beam which satisfies the above requirements \cite{T2K, ichikawa}.
We show the flux of the T2K off-axis $\nu_{\mu}$ beam \cite{ichikawa}
in \figref{fig:flux} (a), for $10^{21}$ POT/yr
at $L$ = 295 km for various off-axis angles between \deg{0} and \deg{3}.
It is clearly seen that the flux peaks at 0.55 to 0.75 GeV at \deg{2} to \deg{3}
off-axis angles.
In \figref{fig:flux}(b) we show the cross section per nucleon
of the $\nu_e$ and $\nu_{\mu}$ CCQE events
off the water target \cite{k2k},
and in \figref{fig:flux}(c),
we show the product of the $\nu_e$ CCQE cross section and the $\nu_{\mu}$
flux at 295 km for various off-axis angles.
Because the neutrino energy reconstruction is essential to determine the 
oscillation phase, we use only the CCQE events in our analysis.
\figref{fig:flux}(b) shows that the CCQE cross sections grow quickly above the threshold,
become $\sim 3.5 \times 10^{-39}$ cm$^2$ at $E_{\nu} \sim 0.6$ GeV,
and stay approximately constant at $\sim 4.5 \times 10^{-39}$
cm$^2$ at $E_{\nu} \simgt 0.8$ GeV up to $\sim 5$ GeV where the flux diminishes.
We also show the typical $\nu_{\mu} \to \nu_e$ transition probability at $L = 295$ km and
that at $L = 1000$ km in \figref{fig:flux}(d), calculated for
$m^2_3 - m^2_1 = 2.5 \times 10^{-3} {\rm eV}^2$,
$m^2_2 - m^2_1 = 8.2 \times 10^{-5} {\rm eV}^2$,
$\sia = 1.0$, $\sis = 0.83$, $\sir = 0.10$,
$\dmns = 0^{\circ}$, and $\rho = 2.8 \density $ for $ L = 295~\km$,
and $\rho = 3.0 \density$ for $L = 1000~\km $.
From \figref{fig:flux} (a), (c), and (d),
we confirm that the \deg{2.0} to \deg{3.0} off-axis beam (OAB) has a strong flux peak
where the oscillation maximum is expected at SK.
We also note that the beam at smaller off-axis angles has significant flux in the
$1.4 \sim 3$ GeV region where the oscillation maximum of the $\nu_{\mu} \to \nu_e$
transition is expected at $L = 1000$ km.
\par
During the T2K experimental period, the center of the $\nu_{\mu}$ beam from
J-PARC goes through Kamioka at \deg{2.0} to \deg{3.0} beneath SK, and the lower
side of the same beam will appear in Korea at various off-axis angles.
\begin{figure}
(a)
\\
\includegraphics[angle = 0 , width = 16cm]{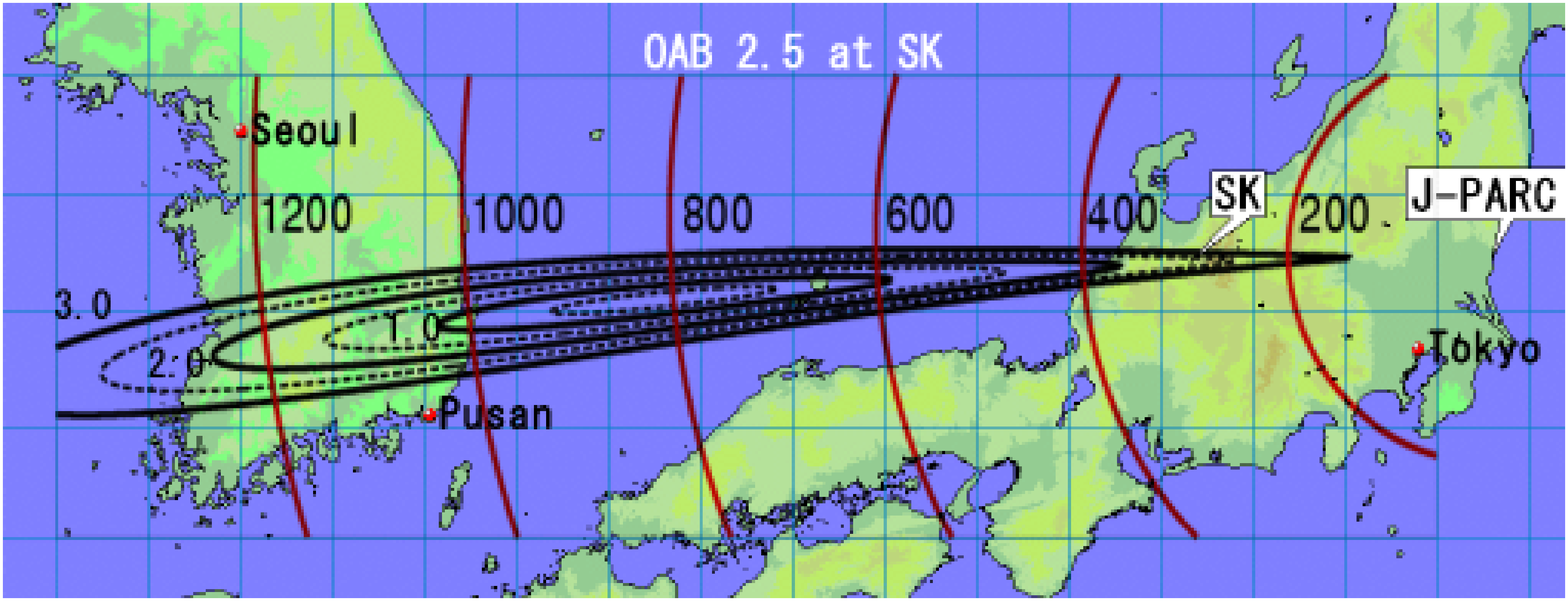}
\\
(b)\\
\includegraphics[angle = 0 , width = 16cm]{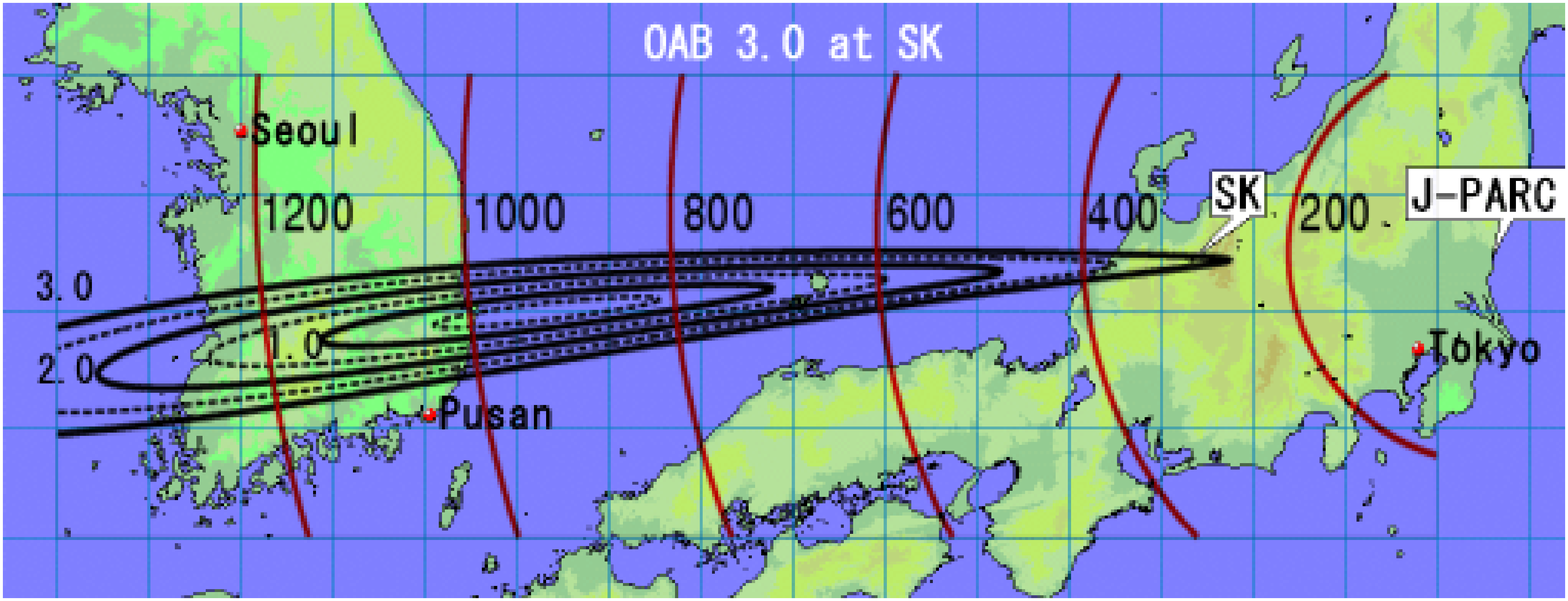}
\caption{The off-axis angle of the neutrino beam from J-PARC on 
the sea level, when the beam center is $2.5^{\circ}$ (a) 
and $3.0^{\circ}$ (b) off at the SK site. 
The baseline lengths are shown by 
vertical contours, and the off-axis angles at the sea level
are shown by elliptic 
contours between $0.5^{\circ}$ and $3.0^{\circ}$.
The SK is slightly off the corresponding contour because it is
about 320 m above sea level. 
}
\label{fig:map}
\end{figure}
We show in \figref{fig:map} the off-axis angle of the $\nu_{\mu}$ beam at
the sea level
for the \deg{2.5} off-axis beam (a) and \deg{3.0} off-axis beam (b) at SK.
The baseline lengths are shown by 
vertical contours, and the off-axis angles at the sea level
are shown by elliptic 
contours between $0.5^{\circ}$ and $3.0^{\circ}$.
The SK is slightly off the corresponding contour because it is
about 320 m above sea level.  
We find that,
when the \deg{2.5} (\deg{3.0}) off-axis beam reaches SK,
the \deg{1.0} (\deg{0.5}) off-axis beam  appears in the
east coast of Korea which is 1000 km away form J-PARC.
If a huge neutrino detector is constructed in the east coast of Korea,
along the T2K beam direction,
the two detector system,
the Tokai-to-Kamioka-and-Korea (T2KK) experiment,
can observe the oscillation maximum of the $\nu_{\mu} \to \nu_e$ transition
probability at two vastly different base-line lengths.
Our most important observation for the T2KK proposal
is that the matter effect term
in \eqref{eq:A-e+} at Korea
is about 3 times as large as that at Kamioka.
Because the sign of the matter effect in $A^e$  follows the sign of
$\Delta_{13}$,
the amplitude of the $\nu_{\mu} \to \nu_e$ oscillation
for the normal hierarchy at $|\Delta_{13}| \sim \pi$
is larger than that for the inverted hierarchy.
Because the matter effect is stronger at Korea,
if the probability is found larger (smaller) at Korea than the one at Kamioka,
we can conclude that the hierarchy is normal (inverted),
irrespective of the other model parameters such as \lsir and \ldmns
\cite{t2kr-l}.
Let us examine this observation semi-quantitatively by using the
approximate formulae in eqs.~(\ref{eq:p-numu-nue}) - (\ref{eq:AB-e}).
The difference between the $\nu_{\mu} \to \nu_e$ oscillation
amplitude 
at a far detector
and that at a near detector is
\begin{equation}
\Delta P_{\nu_\mu \to \nu_e}\left({\Delta_{13}}\right)
= P_{\nu_\mu \to \nu_e}\left(L_{\rm far}; \Delta_{13}\right)
- P_{\nu_\mu \to \nu_e}\left(L_{\rm near}; \Delta_{13}\right) \,.\\
\label{eq:dP}
\end{equation}
%
The error of the $\Delta P_{\nu_\mu \to \nu_e}(\Delta_{13})$
can be estimated as
\begin{eqnarray}
\left[
\delta\left(
\Delta P_{\nu_\mu \to \nu_e}\left(\Delta_{13}\right)
\right)^{^{}}_{_{}}
\right]^2
 &=&
\left[
\delta P_{\nu_\mu \to \nu_e}(L_{\rm near})^{^{}}_{_{}}
{}^{^{}}_{_{}}
\right]^2
+
\left[
\delta P_{\nu_\mu \to \nu_e}(L_{\rm far})^{^{}}_{_{}}
\right]^2
\nonumber \\
&=&
\left(\displaystyle\frac
{P_{\nu_\mu \to \nu_e}(L_{\rm near})}
{\sqrt{N_e^{\rm near}}}
\right)^2
+
\left(
\displaystyle\frac
{P_{\nu_\mu \to \nu_e}(L_{\rm far})}
{\sqrt{N_e^{\rm far}}}
\right)^2
\,.
\label{eq:deltaP}
\end{eqnarray}
%
Here $N_e$ is the number of $\nu_e$ appearance events.
The ratio between $N_e^{\rm far}$ and $N_e^{\rm near}$
at the maximum value of the oscillation probability,
$\Delta_{13}=+\pi$, can be expressed as
\begin{equation}
\displaystyle\frac{N_e^{\rm far}}{N_e^{\rm near}}=
\displaystyle\frac{V_{\rm far}}{V_{\rm near}} 
\,
\displaystyle\frac{
\Phi_{\rm far}
\left(E_{\nu}~{\rm a}t~\Delta_{13} = \pi, L_{\rm far}\right)
}
{
\Phi_{\rm near}
\left(E_{\nu}~{\rm at}~\Delta_{13} = \pi , L_{\rm near}\right)
} \,,
\end{equation}
where $V$ denotes the fiducial volume of the detector
and $\Phi(E_{\nu},~L)$ is the neutrino beam flux at $L$,
which is proportional to $(1/L)^2$.
The cross section ratio at different energies drops out,
because the neutrino cross section of CCQE events is almost 
constant in the 0.7 - 5 GeV region; see \figref{fig:flux}(b).
We therefore need to estimate the number of $\nu_e$ CCQE events
near the oscillation maximum at SK, $N_{\rm near}$.
We show in \figref{fig:event-numbers}, typical numbers of expected
CCQE events for the $\mu$ events (a) and the $e$ events (b),
for the \deg{3.0} OAB at SK.
The open squares show the expected numbers of events in a 200 MeV wide
$E_{\nu}$ bin, after 5 years ($5 \times 10^{21}$ POT),
at $\sir = 0.1$ and $\dmns = 0^{\circ}$ for the normal hierarchy, just
as in \figref{fig:flux} (d).
From the two bins around $E_{\nu} = 0.6$ GeV in \figref{fig:event-numbers}(b),
we may estimate $N_{\rm near} \sim 130$.
\begin{figure}
\begin{center}
\includegraphics[angle = 0 ,width=7.5cm]{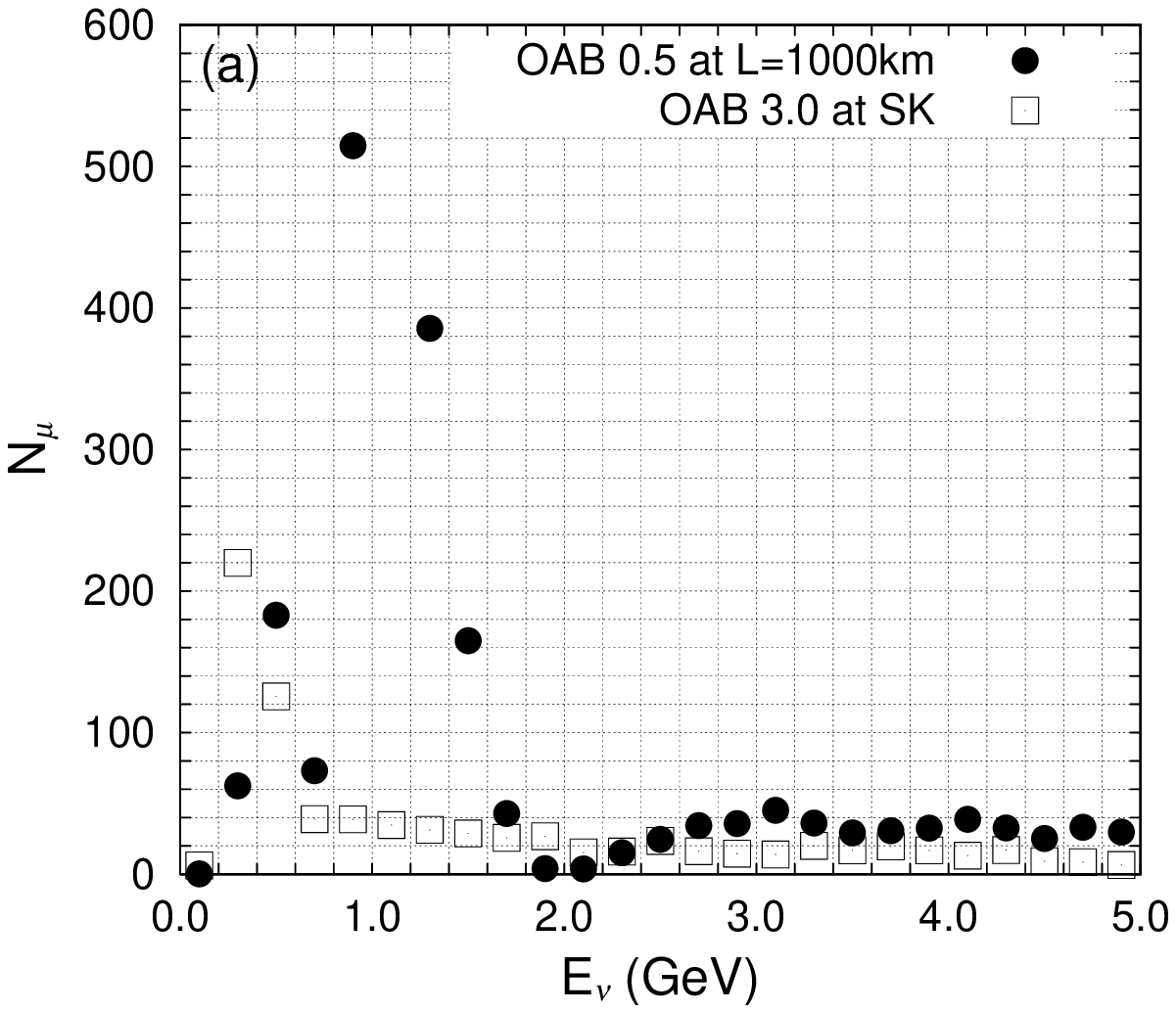}
\includegraphics[angle = 0 ,width=7.5cm]{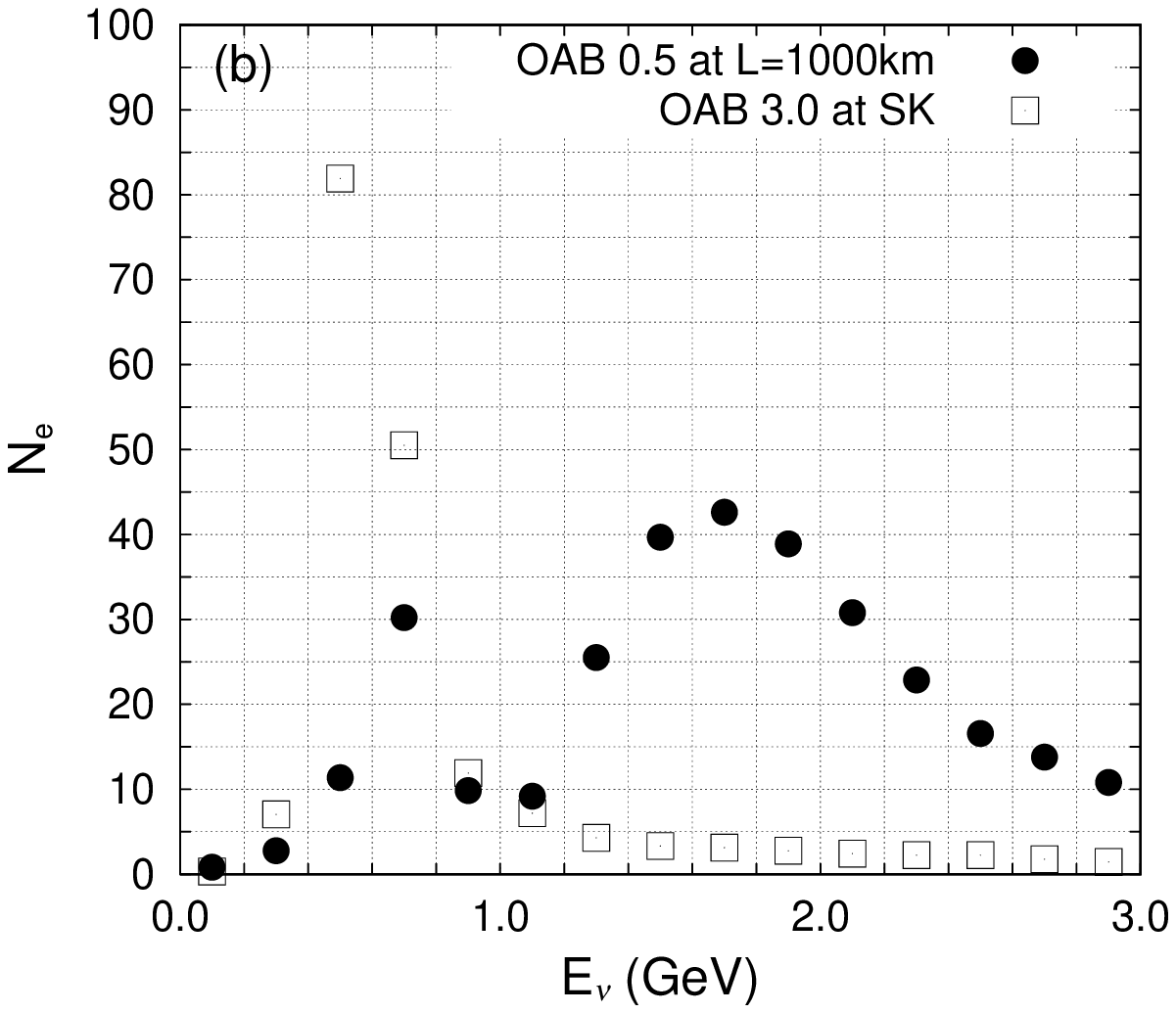}
\end{center}
\caption{The typical numbers of the $\mu$ events (a), and those of the
$e$ events (b),
for the exposure time of 5 years ($5 \times 10^{21}$ POT),
for the $3.0^{\circ}$ OAB at SK (open square), and for the $0.5^{\circ}$ OAB
at $L= 1000~$km with a 100 kt water \cerenkov detector (solid circles).
The input parameters are the same as those of Fig.~\ref{fig:flux}~(d)
\cite{t2kr-l}.
}
\label{fig:event-numbers}
\end{figure}

The difference between the maximum value of the oscillation
probability, $\Delta P_{\nu_\mu \to \nu_e}$, 
for the normal hierarchy ($\Delta_{13} = +\pi$) and
that for the inverted hierarchy ($\Delta_{13} = -\pi$)
can be expressed as
\begin{eqnarray}
\left[
\Delta P_{\nu_\mu \to \nu_e}(+\pi)^{^{}}_{_{}}
-
\Delta P_{\nu_\mu \to \nu_e}(-\pi)^{^{}}_{_{}}
\right]
\sim 0.01
\left(\displaystyle\frac
{\sin^2 2\theta_{\rm \footnotesize RCT}}
{0.10}
\right)
\left(\displaystyle\frac
{L_{\rm far}^{^{}}}
{295 {\rm km}_{_{}}}-1
\right)
\nonumber
\end{eqnarray}
where we set $L_{\rm near} = 295$ km and used the approximation
of eqs.~(\ref{eq:p-numu-nue}) and (\ref{eq:A-e+}).
The difference grows linearly with the distance, $L_{\rm far}$, as
long as the oscillation maximum is covered by the flux.
The significance of excluding the fake hierarchy can then be estimated as
\begin{eqnarray}
\label{eq:significance}
&&\displaystyle\frac
{\Delta P_{\nu_\mu \to \nu_e}\left(+\pi\right)^{}_{}
- \Delta P_{\nu_\mu \to \nu_e}\left(-\pi\right)^{}_{}
}
{\delta^{^{}} ( \Delta P_{\nu_\mu \to \nu_e}(+\pi)^{^{}}_{_{}} )}
\nonumber \\
&&\hspace{5ex}= 2.3
\left( 
          \frac{\sin^2 2\theta_{\rm \footnotesize RCT}}{0.10} 
          \right)^{1/2}
\left( \displaystyle\frac{L_{\rm far}^{^{}}}{295{\rm km}_{_{}}} -1 \right)
\left[
       1+0.225\left(\frac{L_{\rm far}}{295{\rm km}}\right)^{2}
       \frac{100 {\rm kt}}{V_{\rm far}}
\right]^{-1/2}
\,.
\end{eqnarray}
We find that when we put a 100~kt detector at $L=1000$~km,
the significance can exceed the 3-$\sigma$ level in this very rough estimate,
which is confirmed in the following numerical studies.
\par
The phase-shift factor
$B^e$ of \eqref{eq:factorize} also contributes to the determination
of the neutrino mass hierarchy pattern.
Since $\sin^2 (x) $ is an even function of $x$,
the magnitude of $\left | \Delta_{13}/2 + B^e \right |$ controls the
$\nu_{\mu} \to \nu_e$ oscillation in \eqref{eq:factorize}.
Because the value of $\left| \Delta_{13}/2 + B^e \right |$ varies by
changing the sign of $\Delta_{13}$,
we have the possibility to distinguish the sign of $\Delta_{13}$ by
measuring the $\nu_{\mu} \to \nu_e$ oscillation phase.
At Kamioka, however,
$B^e$ is governed by the term proportional to $\cos \dmns$; see
\eqref{eq:B-e+}, and hence the fake hierarchy can reproduce the
same $|\Delta_{13}/2 + B^e  |$ by changing the sign of $\cos \dmns$.
In Korea, the magnitude of the matter effect term in \eqref{eq:B-e+} is 
larger than that of the $\cos \dmns$ terms,
and hence the fake hierarchy cannot reproduce the same oscillation phase.
The efficiency of these phase-shift contribution grows with the
magnitude of $B^e$, which grows as $\cos \dmns$ decreases.
We find that the effect of the phase difference becomes as important
as that of the amplitude difference at $\cos \dmns \sim -1$.
\par
From the above consideration, we observe that it is useful to have a
far detector in Korea in the region where the neutrino flux is
significant around the $\nu_{\mu} \to \nu_e$ oscillation maximum,
which is typically between 1.4 GeV to 3 GeV, see \figref{fig:flux}(d).
Figs.~\ref{fig:flux} (a) and (c) show that beams
at an off-axis angle smaller than \deg{1} have this property.
Figs.~\ref{fig:map}(a) and (b) show that such an off-axis beam will
appear in a specific region near the east coast of Korea during the
T2K experimental period.
\par
Once the neutrino mass hierarchy is determined,
the T2KK experiment
also has the ability to measure
the leptonic CP phase uniquely without using the anti-neutrino beam.
From eqs.~(\ref{eq:p-numu-nue}) and (\ref{eq:A-e+}),
there are 2 unmeasured parameters which control the amplitude of
$\nu_{\mu} \to \nu_e$ oscillation,
$\sin^2 \theta_{\rm \footnotesize RCT}$ and $\sin \dmns$.
Due to the significantly different matter effect at $|\Delta_{13}| =
\pi$ between Kamioka and Korea,
we can constrain both \lsir and \ldmns uniquely from the two amplitudes.
The value of $\cos \dmns$ is measured through the energy dependence of
the $\nu_{\mu} \to \nu_e$
oscillation probability, through the phase shift $B^e$; \eqref{eq:B-e+}.
Since both $\sin \dmns$ and $\cos \dmns$ can be measured independently,
the CP phase \ldmns can be constrained uniquely.
\section{Analysis method}
Before we present the results of our numerical calculation, we would
like to explain our treatment of the signals and background.
In our case study,
we consider a 100~kt level detector,
in order to compensate for the decrease in the neutrino flux 
which is about $(300~{\rm km}/1,000~{\rm km})^2 \sim 1/10$ of that at SK.
We adopt a Water \cerenkov detector because it allows us to distinguish 
clearly the $e^{\pm}$ events
from $\mu^{\pm}$ events.
We use the CCQE events in our analysis,
because it allows us to kinetically reconstruct the neutrino energy
event by event.
Since the Fermi-motion of the target nucleon dominates 
the uncertainty of the neutrino energy reconstruction,
which is about 80 MeV,
we take the width of the energy bin as 
$\delta E_\nu=200$~MeV for $E_\nu > 400$ MeV, in the following analysis
\footnote{At a few GeV region, contributions from soft-charged $\pi$ emission processes to the CCQE
signal events become significant \cite{nakayama}.
Those events have soft charged $\pi$'s which do not emit \cerenkov lights, and hence
the reconstructed neutrino energy is underestimated by about 300 MeV \cite{T2K}.
In this analysis we do not consider contribution from non-CCQE events, and leave
the studies of their impacts for the future.}.
The event numbers of $\nu_\beta$ 
from $\nu_{\alpha}$ flux ($\Phi_{\nu_\alpha}$)
which is delivered by J-PARC \cite{ichikawa}
in the $i$-th energy bin,
$E_\nu^i = 200{\rm MeV} \times (i+1) < E_{\nu} < E_{\nu}^i + \delta E_{\nu}$,
are then calculated as
\begin{equation}
\label{eq:N}
N_\beta^{i} (\nu_\alpha)=
 M N_A
 \int_{E_\nu^i}^{E_\nu^{i}+\delta E_\nu}
 \Phi_{\nu_\alpha}(E)~
 P_{\nu_\alpha \to \nu_\beta}(E)~ 
 \sigma_\beta^{QE}(E)~
 dE\,,
\end{equation}
where
$P_{\nu_{\alpha}\rightarrow \nu_{\beta}}$
is the neutrino oscillation probability 
including the matter effect, 
$M$ is the detector mass,
$N_{A} = 6.017\times10^{23}$ is the Avogadro constant,
and
$\sigma_\alpha^{QE}$ is the CCQE cross section per nucleon in water
\cite{k2k}.
All the primary as well as secondary fluxes used in our analysis are
obtained from the website \cite{ichikawa}.
The fiducial volume of Super-Kamiokande is 22.5 kt,
and we assume that a detector in Korea is 100~kt.
CCQE events have been selected as single-ring events at K2K. Efficiency of
the technique has been estimated to be about $80 \%$ around 600 MeV and
about $60 \%$ around a few GeV \cite{T2K, nakayama}.
In the following numerical analysis
we set the efficiency to be $100 \%$ at both SK and at a far detector
for brevity
\footnote{
Our results can be regarded as those for larger fiducial volumes,
such as 28.1 kt at SK and 167kt at a far detector, if we take the efficiencies
of $80\%$ and $60\%$ estimated in refs \cite{T2K, nakayama}.
}.
\par
When the secondary neutrino flux 
($\nu_e\,,\bar{\nu}_e\,,\bar{\nu}_{\mu}$)
of the  $\nu_\mu$ primary beam is considered,
the $e$-like and $\mu$-like events for the $i$-th bin are obtained
as 
\begin{equation}
\label{signal}
 N_{\alpha}^{i} = N_{\alpha}^{i}(\nu_\mu) 
+\sum_{\nu_\beta = \nu_e\,,\bar{\nu}_e\,,\bar{\nu}_{\mu}} 
N_{\alpha}^{i}(\nu_\beta) 
\,,
{\mbox{\hspace*{5ex}}}
(\alpha = e\,, \mu)\,.
\end{equation}
The second term in eq.(\ref{signal})
corresponds to the contribution from the secondary
neutrino flux.
In our analysis,
we calculate $N^i_{\mu,e}$ 
by assuming the
following input parameters:
\begin{equation}
\left.
\begin{array}{l}
|m^2_3 - m^2_1 |^{\rm input} = 2.5 \times 10^{-3}~\mbox{eV}^2 \,,\\
(m^2_2 - m^2_1)^{\rm input}  = 8.2 \times 10^{-5}~\mbox{eV}^2\,, \\
\sin^2 \theta_{\rm \footnotesize ATM}^{\rm input} = 0.5 \,,  \\
\sin^2 2\theta_{\rm \footnotesize SOL}^{\rm input} = 0.83 \,, \\ 
\end{array}
\right\}
\label{eq:input}
\end{equation}
with the constant matter density, $\rho^{\rm input}=2.8~{\rm g/cm}^3$ along T2K
and $\rho^{\rm input}=3.0~{\rm g/cm}^3$ for the Tokai-to-Korea baseline,
which goes through deeper in the earth than that of T2K\footnote{{%
Because the sea between Japan and Korea is less than 1km deep,
there is no contribution from the water to the average density.}}.
We use the above numerical values in order to keep the consistency
with our previous report \cite{t2kr-l}.
Our results are not sensitive to the small change in the input values
in eq.(\ref{eq:input})
but the quantitative results on the mass hierarchy determination
depend on the mater density along the baseline.
Dedicated study on the impacts of the matter density profile along the
T2KK baseline will be reported elsewhere.
We  examine various input values of \lsir, \ldmns,
and the sign of $m_3^2 - m_1^2$.
Since our concern is the possibility to distinguish the neutrino mass 
hierarchy and to constrain the CP phase uniquely,
we study how the above `data' can constrain the model parameters by
using the $\chi^2$ function
\begin{equation}
\label{chi^2 define}
\Delta\chi^2 = \chi^2_{\rm SK} + \chi^2_{\rm Kr} + \chi^2_{\rm sys} 
+ \chi^2_{\rm para}\,.
\end{equation}
Here the first two terms, $\chi^2_{\rm SK}$ and $\chi^2_{\rm Kr}$,
measure the parameter dependence of the fit to the SK and the Korean 
detector data,
\begin{eqnarray}
\label{eq:chi^2event}
 \chi^2_{\rm SK,Kr}
 = \sum_{i} \sum_{\alpha = e,\mu}
\left(
\displaystyle\frac
{(N_\alpha^{i})^{\rm fit} - N_\alpha^{i}}
{\sqrt{N^i_\alpha}}
\right)^2
\,,
\end{eqnarray}
where the summation is over all bins
from 0.4 GeV to 5.0 GeV for $N_{\mu}$,
0.4 GeV to 1.2 GeV for $N_{e}~$at SK,
and 0.4 GeV to 2.8GeV for $N_{e}~$at Korea.
Here $N^i_{\mu,e}$ is the calculated number of events
in the $i$-th bin,
and its square root gives the statistical error.
We include the contribution of the $\mu$-events in order to constrain
the absolute value of $\Delta_{13}$  strongly in this analysis,
because a small error of $\Delta_{13}$ dilutes the phase shift $B^e$
\cite{t2kr-l, koike05}.

$N_{i}^{\rm fit}$ is calculated by allowing the model parameters
to vary freely
and by allowing for systematic errors.
In our analysis, we consider 4 types of systematic errors.
The first ones are for the overall normalization of each neutrino flux,
for which we assign $3\%$ errors,
\begin{equation}
f_{\nu_\beta} = 1 \pm 0.03\,,
\end{equation}
for all neutrino flavor,
which are taken common for T2K and the Tokai-to-Korea experiment.
The second systematic error is for the uncertainty in the matter density,
for which we allow $3\%$ overall uncertainty along the baseline, 
independently for T2K ($f^{\rm SK}_{\rho}$) and
the Tokai-to-Korea experiment ($f^{\rm Kr}_{\rho}$):
\begin{equation}
\rho_{i}^{\rm fit} = f^{i}_{\rho}\,\rho^{\rm input}_i
\hspace{5ex}
(i = {\rm SK,~Kr}) \,.
\end{equation}
The third uncertainty is for the CCQE cross section.
Since $\nu_e$ and $\nu_\mu$ CCQE cross sections are expected to be very 
similar theoretically, we assign a common overall error of $3\%$ for 
$\nu_e$ and $\nu_{\mu}$ 
($f_e^{_{\rm QE}} = f_\mu^{_{\rm QE}} \equiv f_\ell^{_{\rm QE}}$), 
and an independent $3\%$ error for $\bar{\nu}_e$ and $\bar{\nu}_\mu$
CCQE cross sections 
($f_{\bar{e}}^{_{\rm QE}} = f_{\bar{\mu}}^{_{\rm QE}}
 \equiv f_{\bar{\ell}}^{_{\rm QE}}$).  
The last one is the uncertainty of the fiducial volume,
for which we assign $3\%$ error independently for T2K 
($f_{\rm V}^{_{\rm SK}}$) and the
Tokai-to-Korea experiment ($f_{\rm V}^{_{\rm Kr}}$).
$N_{\alpha}^{i,{\rm fit}}$ is then calculated as
\begin{eqnarray}
\left[ N_\alpha^{i,{\rm fit}}(\nu_\beta) \right]_{\rm at~SK, Kr}&=&
f_{\nu_\beta}\, f_{\alpha}^{_{\rm QE}}\, f_{\rm V}^{_{\rm SK,Kr}}\,
N_\alpha^{i}(\nu_\beta)\,, 
\end{eqnarray}
and $\chi^2_{\rm sys}$ has four terms;
\begin{equation}
\label{chisq-sys}
 \chi^2_{\rm sys} = 
\sum_{\alpha = e,\bar{e},\mu,\bar{\mu}}
\left(
\displaystyle\frac{f_{\nu_{\alpha}}-1}{0.03}
\right)^2
+
\sum_{\alpha = l, \bar{l}}
\left(
\displaystyle\frac{f^{\rm QE}_{\alpha}-1}{0.03}
\right)^2
+
\sum_{ i = {\rm SK,~Kr}}
\left\{
\left(
\displaystyle\frac{f^{i}_{\rho}-1}{0.03}
\right)^2
+
\left(
\displaystyle\frac{f^{i}_{\rm V}-1}{0.03}
\right)^2
\right\}\,.
\end{equation}
In short,
we assign $3\%$ errors for the normalization of each neutrino flux,
the $\nu_l$ and $\bar{\nu}_l$ CCQE cross sections,
the effective matter density along each base line,
and for the fiducial volume of SK, and that of the Korean detector.
\par
Finally, $\chi^2_{\rm para}$ accounts for the external constraints 
on the model parameters:
\begin{eqnarray}
\label{eq:chisq-para}
\chi^2_{\rm para}
&=&
\left(
\displaystyle\frac{(m_2^2 - m_1^2)^{\rm fit} - (m_2^2 - m_1^2)^{\rm input}}
{ 0.6 \times 10^{-5}}
\right)^2
+
\left(
\displaystyle\frac
{\sin^22\theta_{\rm \footnotesize SOL}^{\rm fit}- \sin^22\theta_{\rm \footnotesize SOL}^{\rm input}}
{0.07}
\right)^2 
\nonumber
\\
&&
+
\left(
\displaystyle\frac
{\sin^22\theta_{\rm \footnotesize RCT}^{\rm fit}- \sin^22\theta_{\rm
\footnotesize RCT}^{\rm input}}
{0.01}
\right)^2\,.
\end{eqnarray}
The first two terms correspond to the present experimental constraints
summarized in eq.~(\ref{eq:sol-data}).
In the second term,
we adapt the larger error than that shown in \eqref{eq:sol-angle}
for continuity of the previous work, ref.~\cite{t2kr-l}.
In the last term, we assume that the planned future reactor
experiments \cite{double-chooz, kaska}
should measure \lsir with the expected uncertainty of 0.01, during the
T2KK experimental period.
In total, our $\Delta\chi^2$ function depends on 16 parameters,
the 6 model parameters and the 10 normalization factors.
%
\section{Determination of the mass hierarchy pattern}
 In this section we show the results of our numerical calculation. 
First, we search for the best combination of the off-axis angle at SK
and that at a Korean detector,
as well as
the base-line length up to Korean detector for determining the sign of
$m^2_3 - m^2_1$.
\begin{figure}[t]
\begin{center}
\includegraphics[angle = 0 ,width=7.5cm]{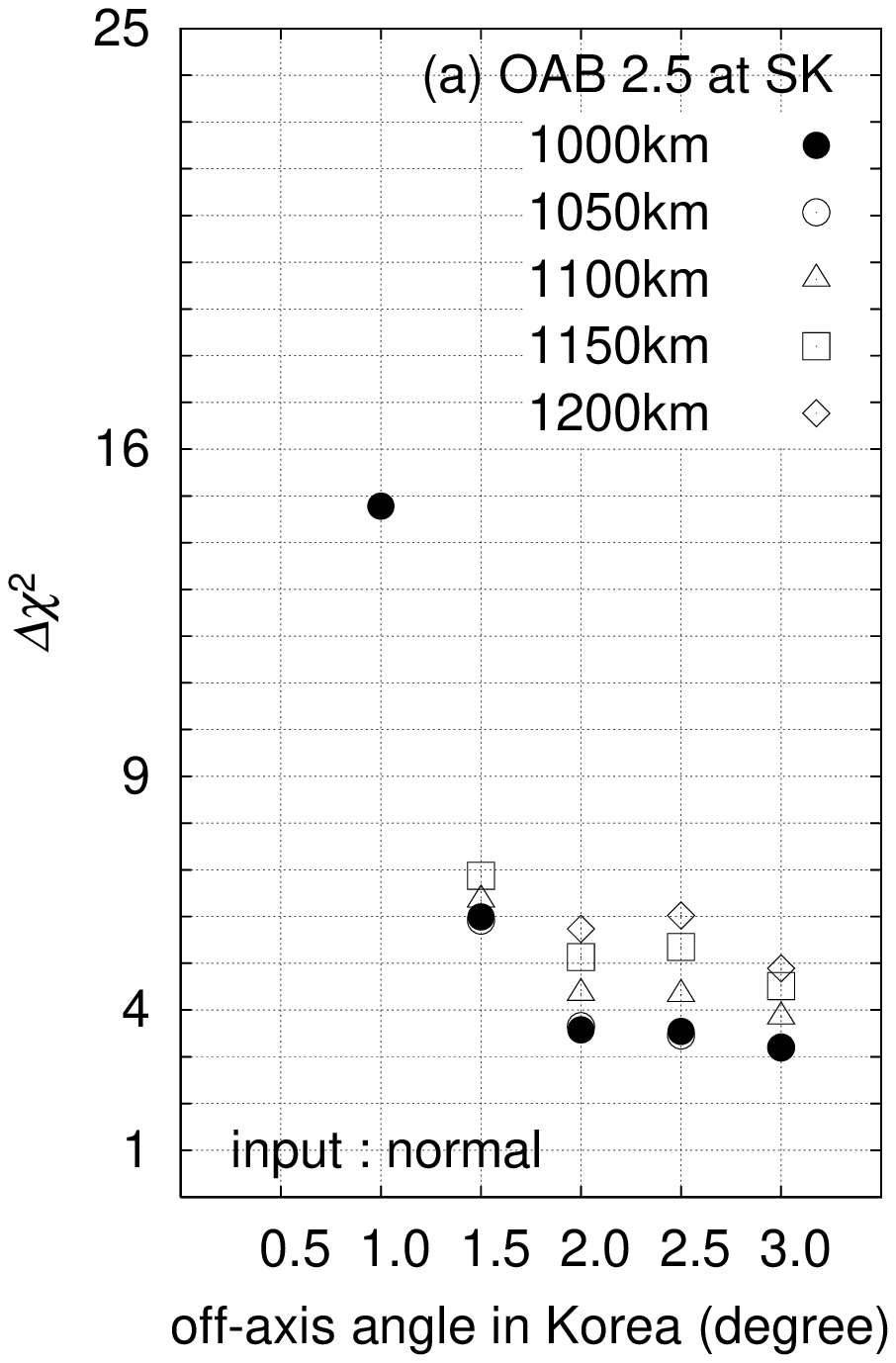}
\includegraphics[angle = 0 ,width=7.5cm]{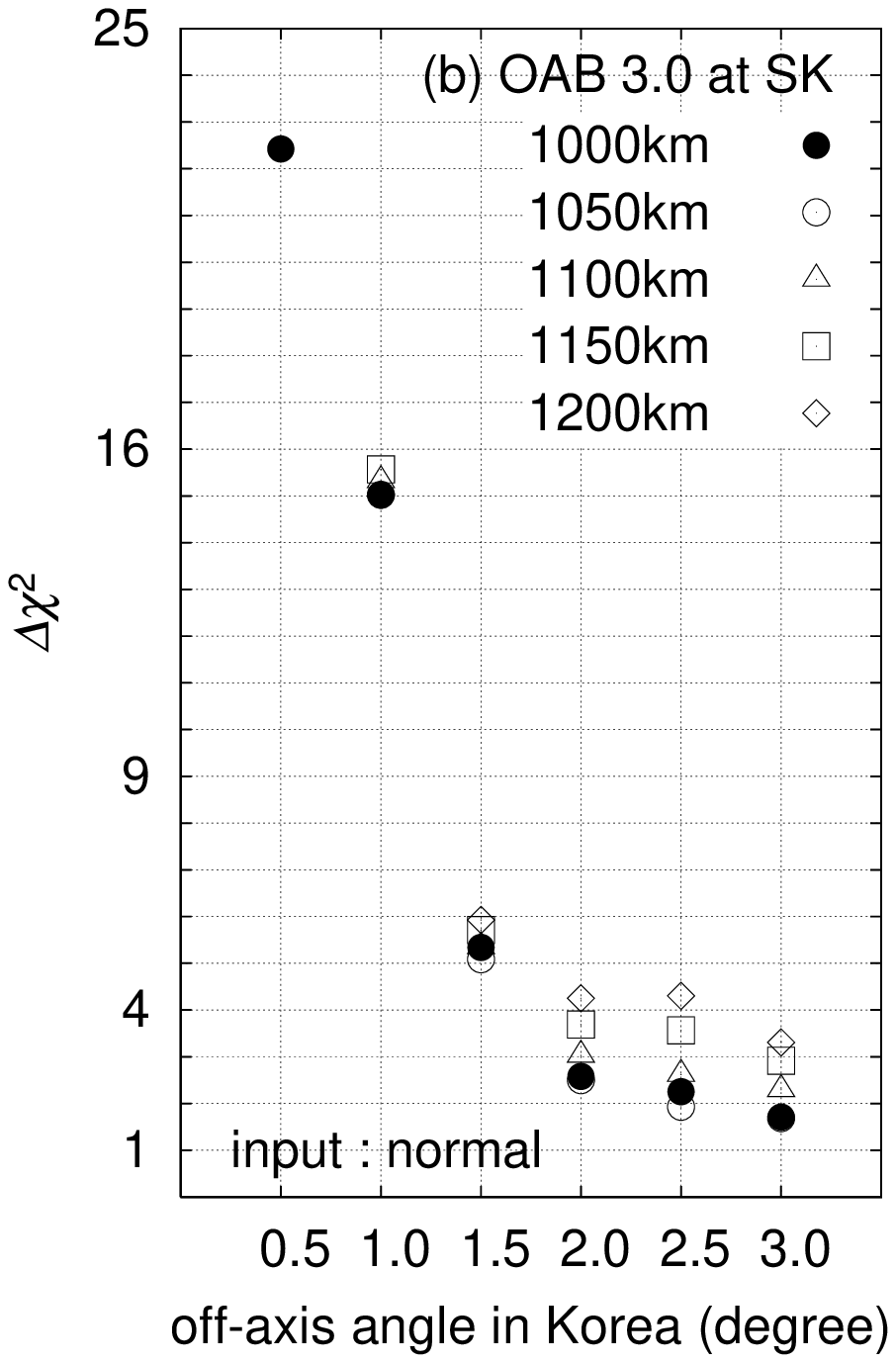}
\end{center}
\caption{Minimum $\Delta\chi^2$ of the T2KK two detector experiment
after 5 years of running ($5 \times 10^{21}$ POT)
as functions of the off-axis angle and
the base-line length of the far-detector from J-PARC at Tokai,
when the normal hierarchy 
($m^2_3 - m^2_1 > 0$)
is assumed in generating the events,
and the inverted hierarchy 
($m^2_3 - m^2_1 < 0$) is assumed in the fit. 
The left-hand figure (a) is for the $2.5^\circ$ OAB at SK,
and the right-hand one (b) is for the $3.0^{\circ}$ OAB beam at SK.
The input parameters are the same as those of Fig.~\ref{fig:flux}~(d);
in particular,
$\sin^2 2\theta_{\rm \footnotesize RCT}^{\rm input} = 0.10$ and $
 \dmns^{\rm input} = 0^{\circ}$.
}
\label{fig:place-n}
\end{figure}
For this purpose, we first calculate the expected number of the
$\nu_{\mu} \to \nu_e$ CCQE events
at both detectors by assuming either normal or inverted hierarchy, and
then examine if the resulting `data' can be fitted for the opposite
hierarchy by adjusting the model parameters.

We show in \figref{fig:place-n} 
the minimum $\Delta\chi^2$ expected at the T2KK two detector experiment
after 5 years of running ($5 \times 10^{21}$POT),
as functions of the off-axis angle and
the base-line length of the far-detector site from J-PARC at Tokai,
when the normal hierarchy 
($m^2_3 - m^2_1 > 0$)
is assumed in generating the events,
and the inverted hierarchy 
($m^2_3 - m^2_1 < 0$) is assumed in the fit. 
The left-hand figure (a) is for the $2.5^\circ$ OAB at SK,
and the right-hand one (b) is for the $3.0^{\circ}$ OAB beam at SK.
The input parameters are choose as in \eqref{eq:input},
$\sin^2 2\theta_{\rm \footnotesize RCT}^{\rm input} = 0.10$ and $
\dmns^{\rm input} = 0^{\circ}$.
The four symbols, solid circle, open circle, triangle, and square are 
for $ L = 1000$km, 1050km, 1100km, and 1150km, respectively.
There are no data points at \deg{0.5} in \figref{fig:place-n}(a) for
the \deg{2.5}
OAB at SK,
because the
\deg{0.5} off-axis beam does not reach Korea: see \figref{fig:map}(a).
It is clearly seen from \figref{fig:place-n} that the best combination
of off-axis angles
are \deg{3} for SK and \deg{0.5} for the Korean detector at $L = 1000$ km.
The \deg{0.5} off-axis beam has strong flux up to $\sim 2.2$ GeV,
which overlaps significantly with the broad peak of the $\nu_{\mu} \to \nu_e$
oscillation at $L = 1000$ km; see \figref{fig:flux} (a), (c) and (d).
Because the number of the $\nu_e$ CCQE events
is large enough around the oscillation maximum for $\sir \sim 0.1$,
both at SK and at the far detector in Korea, we are able to measure
the difference in the magnitude
of the $\nu_{\mu} \to \nu_e$ probability at two vastly different
baselines, and can hence distinguish between the normal
hierarchy and the inverted hierarchy.
We can reject the fake hierarchy
at 4.7-$\sigma$ level in our simple simulation with this combination
of \deg{3.0} at SK and \deg{0.5} at
$L = 1000$ km.
If we remove the the constraint of the future reactor
experiment, the last term in \eqref{eq:chisq-para},
the minimum $\Delta \chi^2$ value drops from 23 to 18,
for the combination of 
\deg{3.0} at SK and \deg{0.5} at $L = 1000$ km;
see \cite{t2kr-l}.
\par
We find from \figref{fig:place-n}
that the \deg{1.0} OAB in Korea still keeps the sensitivity to the
neutrino mass hierarchy,
where both a combination of \deg{1.0} at $L = 1000$ km and \deg{2.5}
OAB at SK (\figref{fig:place-n} (a)) 
and that of \deg{1.0} at $L = 1000$ km $\sim 1150$ km and the
\deg{3.0} OAB at SK
(\figref{fig:place-n} (b))  distinguish the neutrino mass hierarchy
nearly at 4-$\sigma$ level in our simulation.
This is because the CCQE cross section times the flux of \deg{1.0} OAB
extends to $\sim$ 1.7 GeV,
see \figref{fig:flux} (a) and (c),
which barely overlaps with the broad peak region of the $\nu_{\mu} \to \nu_e$ oscillation probability
shown in \figref{fig:flux}(d).
From \figref{fig:map} (a) and (b),
we find that the \deg{1.0} OAB is observable only in the east coast of
Korea ($L \sim 1000$ km)
for the \deg{2.5} OAB at SK,
whereas for the \deg{3.0} OAB at SK,
it can be observed at various base-line lengths up to $\sim 1150$ km.
The small values of $\Delta \chi^2$ for larger off-axis angles in \figref{fig:place-n}
tell us that it is essential to choose the location of the detector in Korea where the off-axis angle is smaller than \deg{1.0}.
\begin{figure}[t]
\begin{center}
\includegraphics[angle = 0 ,width=7.5cm]{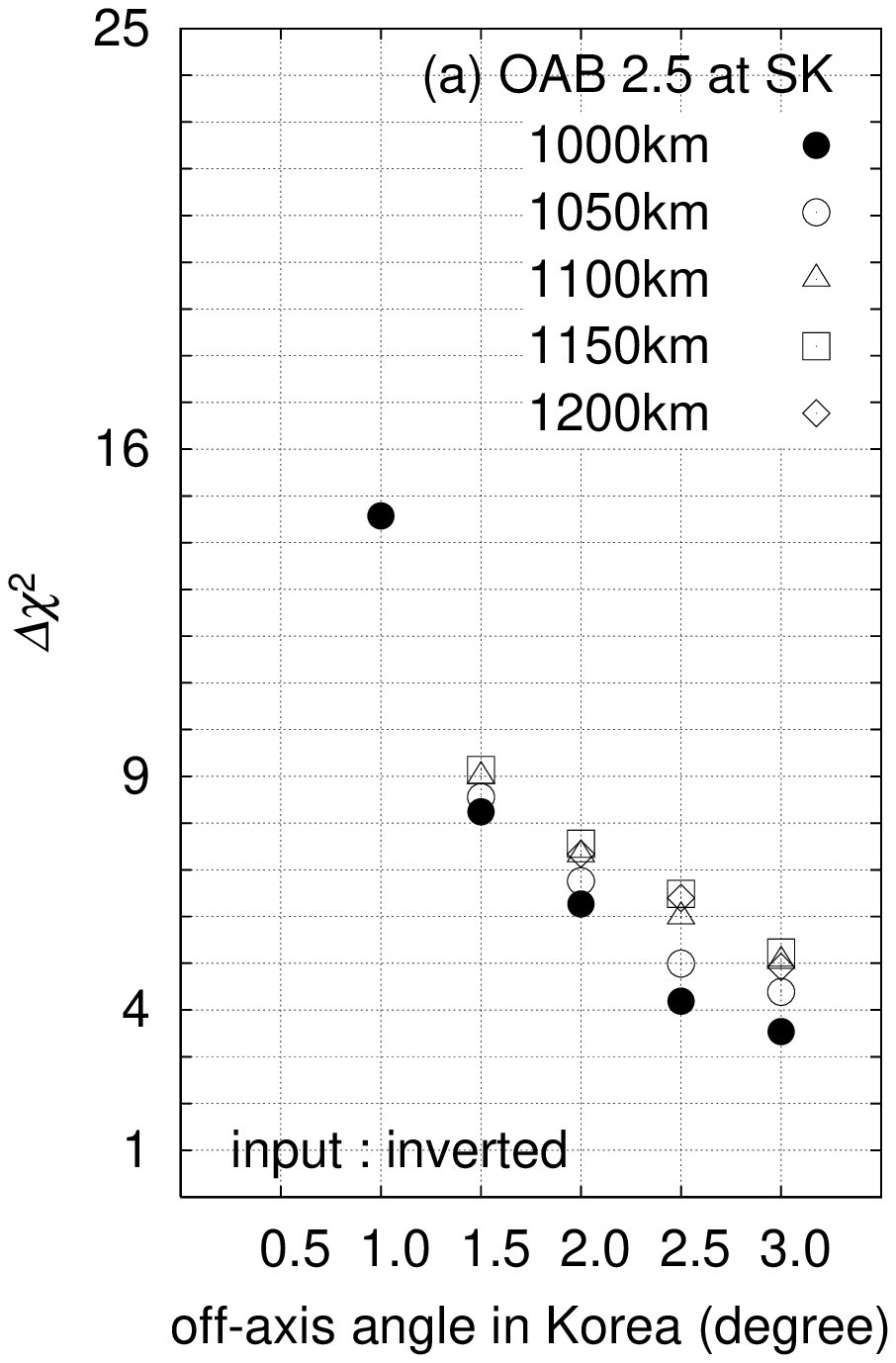}
\includegraphics[angle = 0 ,width=7.5cm]{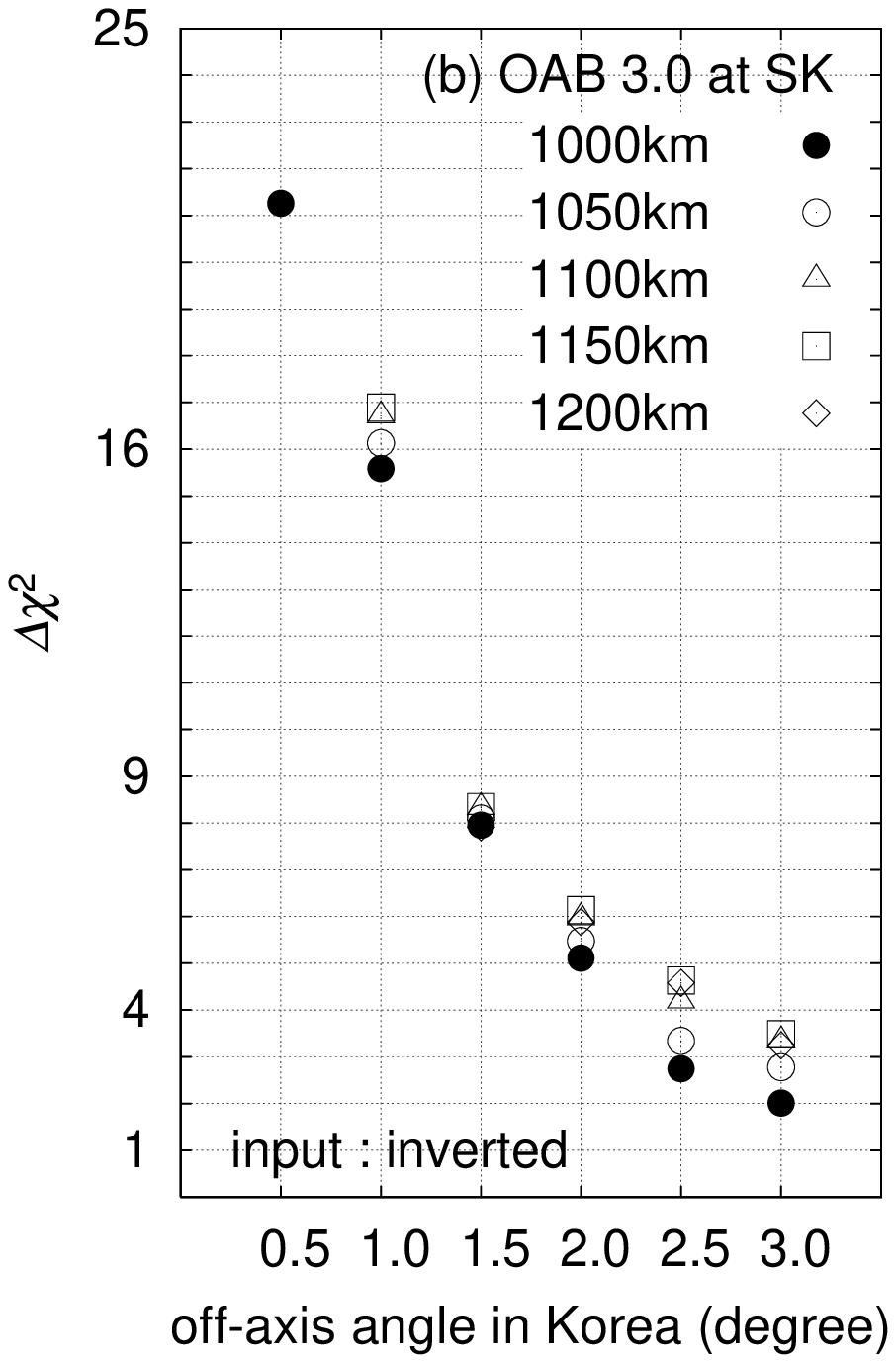}
\end{center}
\caption{ The same as \figref{fig:place-n},
but when the input data is calculated for the inverted hierarchy ($m^2_3 - m^2_1 < 0$)
and the fit is performed by assuming the normal hierarchy ($m^2_3 - m^2_1 > 0$).
All the other parameters are the same as those in \figref{fig:place-n}.
}
\label{fig:place-i}
\end{figure}
\par
\figref{fig:place-i} is the same as \figref{fig:place-n}, but when the input data is calculated for the inverted hierarchy
and the fit is performed by assuming the normal hierarchy.
It is remarkable that almost the same level of the capability to distinguish the neutrino mass hierarchy
can be achieved for the OAB $\simlt 1^{\circ}$ at $L \sim 1000$ km
even when the hierarchy is inverted ($m_3^2 - m_1^2< 0$).
Slight decreases of the minimum $\Delta \chi^2$ value of the best combinations,
23.5 in \figref{fig:place-n}(b) to 21.3 in \figref{fig:place-i}(b), and
14.8 in \figref{fig:place-n}(a) to 14.6 in \figref{fig:place-i}(a),
can be attributed to the smaller expected number of the $\nu_e$
appearance events because of the matter effect which suppresses the probability in the inverted hierarchy.
We may conclude that a far detector that observes the T2K neutrino beam at $L \sim 1000$ km
and the off-axis angle $\simlt$\deg{1.0}  has the potential to determine the neutrino mass hierarchy whether it is normal
or inverted.
\par
Because we find from \figref{fig:place-n} and \figref{fig:place-i}
that a combination of \deg{0.5} OAB at $L \sim 1000$ km and the \deg{3.0} OAB at SK has a significantly
better capability of determining the neutrino mass hierarchy,
we study in the following physics potential of this preferred T2KK set up in more detail.
In particular,
we investigate
the whole un-explored parameter space of the three neutrino model.
\par
\begin{figure}[t]
\begin{center}
\includegraphics[angle = 0 ,width=7.8cm]{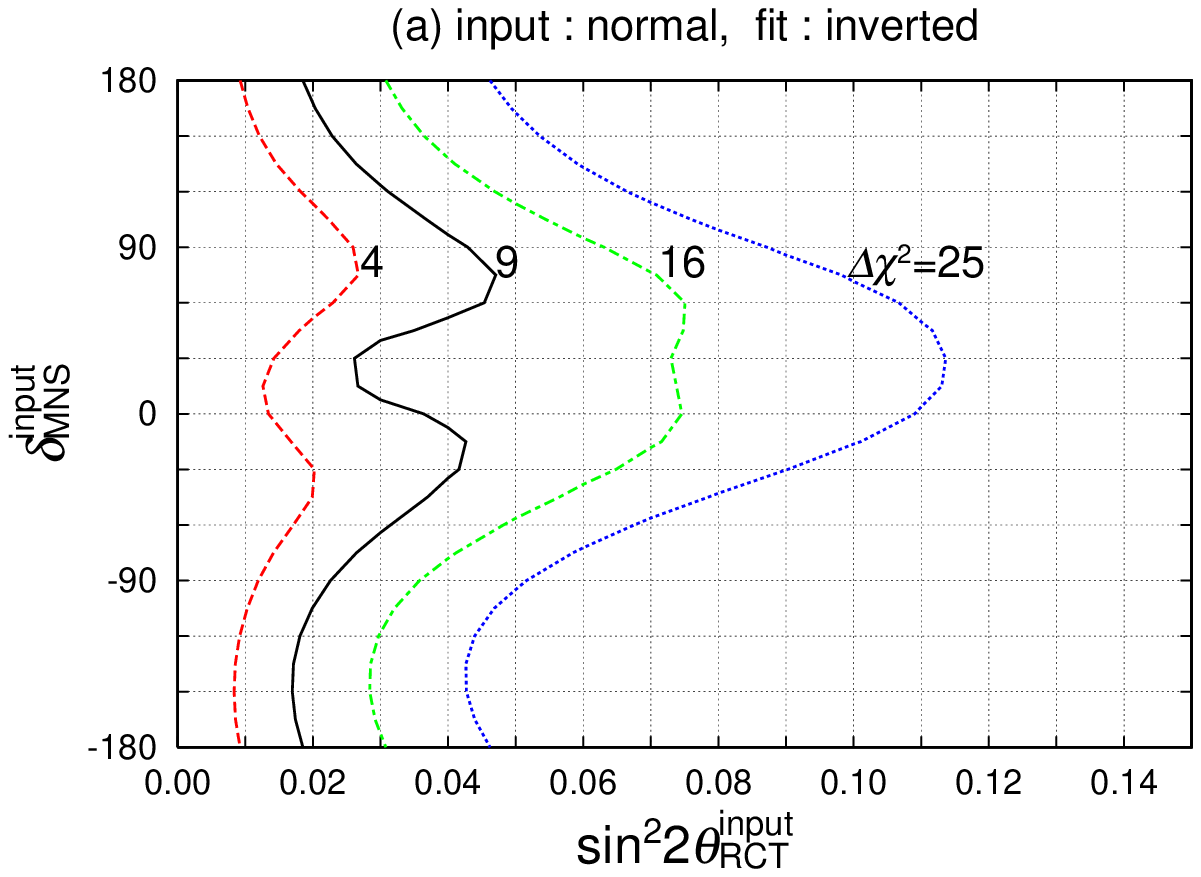}
\includegraphics[angle = 0 ,width=7.8cm]{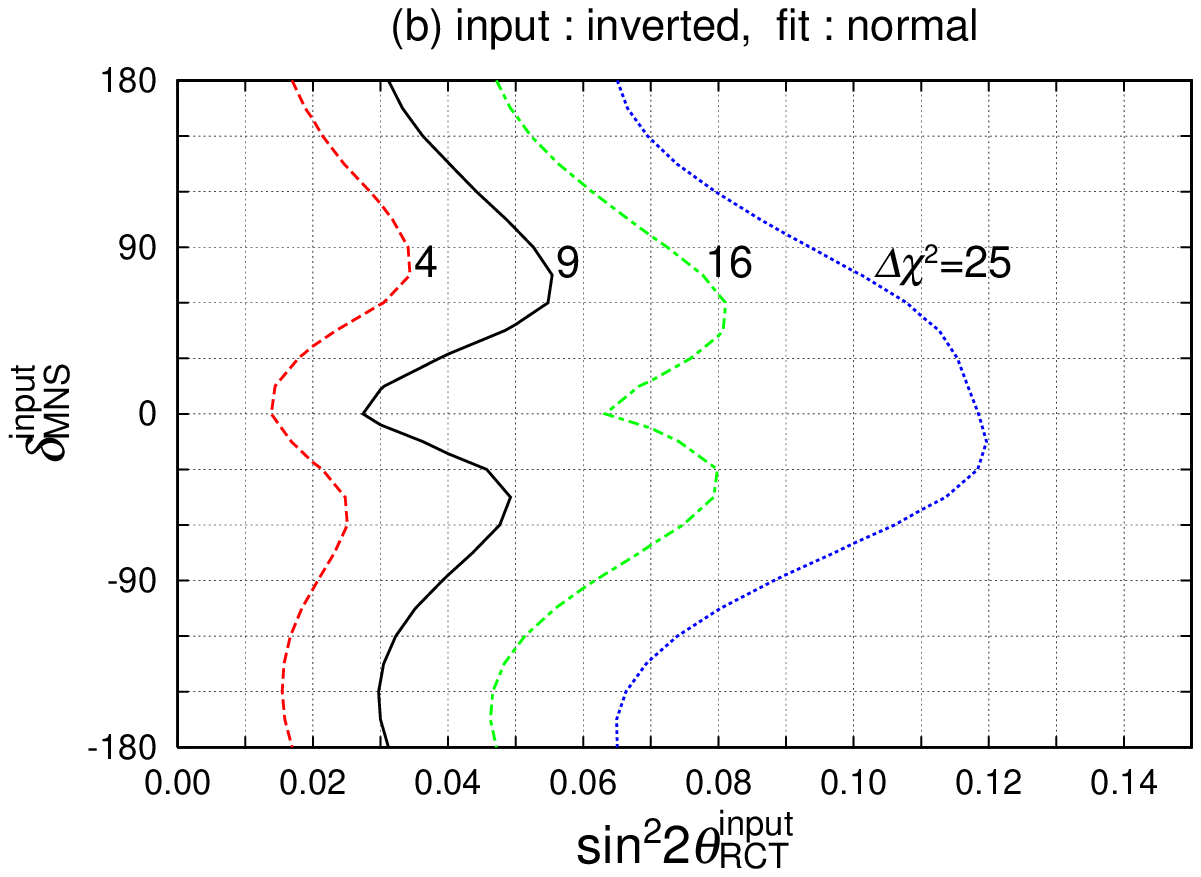}
\end{center}
\caption{
Capability of the T2KK two-detector experiment to determine the neutrino mass hierarchy,
(a) when the mass hierarchy is normal $(m^2_3 - m^2_1 > 0)$,
and (b) when it is inverted $(m^2_3 - m^2_1 < 0)$.
The numerical results are obtained for a combination of \deg{3.0}OAB at SK
and \deg{0.5}off-axis at $L = 1000 \km$ with a 100 kt water \cerenkov detector,
after 5 years of running ($5 \times 10^{21}$ POT).
In each figure the input data are calculated for the model parameters
at various $\sir^{\rm input}$ and $\dmns^{\rm input}$,
and the fit has been performed by surveying the whole parameter space with the
opposite mass hierarchy.
The resulting values of minimum $\Delta \chi^2$ are shown as contours for 2, 3, 4, and 5$\sigma$.
The wrong hierarchy can be excluded with the corresponding confidence level
if the true \lsir and \ldmns values lie in the right-hand side of each contour.
The model parameters are set at
$(m^2_3 - m^2_1)^{\rm input} = 2.5 \times 10^{-3} {\rm eV}^2$ (a),
$-2.5 \times 10^{-3} {\rm eV}^2$ (b),
$(m^2_2 - m^2_1)^{\rm input} = 8.2 \times 10^{-5} {\rm eV}^2$,
$\sia^{\rm input} = 1.0$,
$\sis^{\rm input} = 0.83$,
$\rho^{\rm input} = 2.8 \density $ for SK,
and $\rho^{\rm input} = 3.0 \density$ for $L = 1000 \km$.
}
\label{fig:sensitivity}
\end{figure}

First in \figref{fig:sensitivity}, we summarize our findings on the capability
of the T2KK experiment to determine the neutrino mass hierarchy in the
whole space of \lsir and \ldmns.
\figref{fig:sensitivity}(a) shows our result when the mass hierarchy is normal ($m_3^2 - m_1^2 > 0$),
and (b) when it is inverted ($m_3^2 - m_1^2 < 0$).
In each figure the input data are calculated for the model parameters
at various $\sir^{\rm input}$ and $\dmns^{\rm input}$ points,
and the fit has been performed by surveying the whole parameter space, but under
the opposite mass hierarchy.
The resulting values of the minimum $\Delta \chi^2$ are shown as contours for 2, 3, 4, and 5-$\sigma$.
The wrong mass hierarchy can be excluded with the corresponding confidence level
if the true \lsir value lies in the right-hand side of each contour along the true value of \ldmns ($\dmns^{\rm input}$).
In particular, the minimum $\Delta \chi^2$ values of 22 for the point ($\sir^{\rm input},~\dmns^{\rm input}$)
= (0.10, \deg{0}) in \figref{fig:sensitivity}(a)
corresponds to the highest point in \figref{fig:place-n}(b),
and the corresponding value of 21 in \figref{fig:sensitivity}(b) is the highest point in \figref{fig:place-i}(b).
We find that the wrong hierarchy can be excluded at the 3-$\sigma$ level if $\sir^{\rm input} > 0.05$ (0.06)
if the hierarchy is normal (inverted).
\par
It is remarkable that the \ldmns = \deg{0} case chosen to plot \figref{fig:place-n} and \figref{fig:place-i}
turns out to be the case when it is most difficult to determine the neutrino mass hierarchy.
If \ldmns = \deg{180},
the wrong hierarchy can be excluded at the 3-$\sigma$ level
for $\sir^{\rm input} \simgt 0.02$ for the normal hierarchy
(\figref{fig:sensitivity}(a)) or $\sir^{\rm input} \simgt 0.03$
for the inverted hierarchy (\figref{fig:place-i}(b)).
The origin of the \ldmns dependence is the difference of the oscillation phase at
the far detector in Korea.
From eqs.~(\ref{eq:B-e+}) and (\ref{eq:factorize}), the difference of the oscillation
phase near the oscillation maximum, $|\Delta_{13}| = \pi$,
between the input and the fit is expressed as,
\begin{eqnarray}
&&
\left | \frac{\Delta_{13}}{2} + B^e_{\rm input} \right| -
\left | - \frac{\Delta_{13}}{2} + B^e_{\rm fit} \right |
\nonumber \\
&\sim &
 \pm 0.15 \left  \{ 
\cos \dmns^{\rm input} \left ( \frac{  0.10 } {\sin^2 2\theta_{\rm \footnotesize RCT}^{\rm input} } \right )^{1/2}
+ \cos \dmns^{\rm fit} \left ( \frac{  0.10 } {\sin^2 2\theta_{\rm \footnotesize RCT}^{\rm fit} } \right )^{1/2}
\right \}
\left ( \frac{  | \Delta_{13} |} {\pi} \right ) 
\nonumber
\\ &&\mp 0.58 \left ( \frac{L}{1000 \rm{km}} \right )\,.
\label{eq:Dphase}
\end{eqnarray}
The upper sign is for the normal hierarchy, and
the lower sign is for the inverted hierarchy.
The phase-shift difference depends on both $\cos \dmns^{\rm input}$
and $\cos \dmns^{\rm fit}$.
As explained in section 3, below \eqref{eq:significance},
when $\cos \dmns^{\rm input} \sim 1$ ($\dmns^{\rm input} \sim 0^{\circ}$)
the phase shift is smaller than that with the other 
$\dmns^{\rm input}$ at $L=1000$ km.
Therefore the fitted value
of $\cos \dmns$ ($\cos \dmns^{\rm fit}$)
also tends to be large and has the opposite sign for the fake hierarchy.
If $\cos \dmns^{\rm input} \sim -1$, it is not possible to compensate
for the phase-shift difference of \eqref{eq:Dphase}
even by choosing $\cos \dmns^{\rm fit} = 1$,
and the significantly higher minimum $\Delta \chi^2$ value results in \figref{fig:sensitivity}(a) and (b).
In general, $\cos \dmns^{\rm fit} > 0$ is favored even when
$\cos \dmns^{\rm input} < 0$  in order to minimize the phase-shift difference of \eqref{eq:Dphase}.
\begin{figure}
\begin{center}
\includegraphics[angle = 0 ,width=7.8cm]{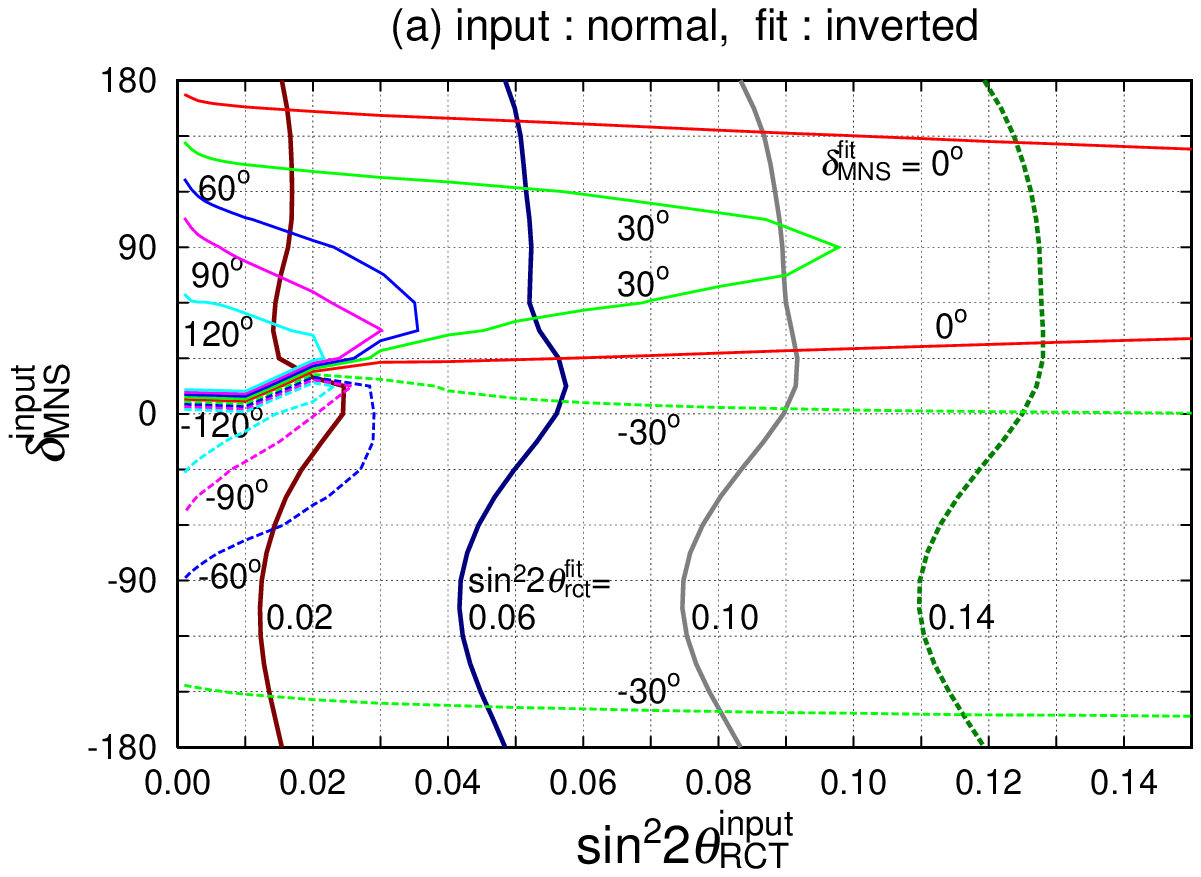}
\includegraphics[angle = 0 ,width=7.8cm]{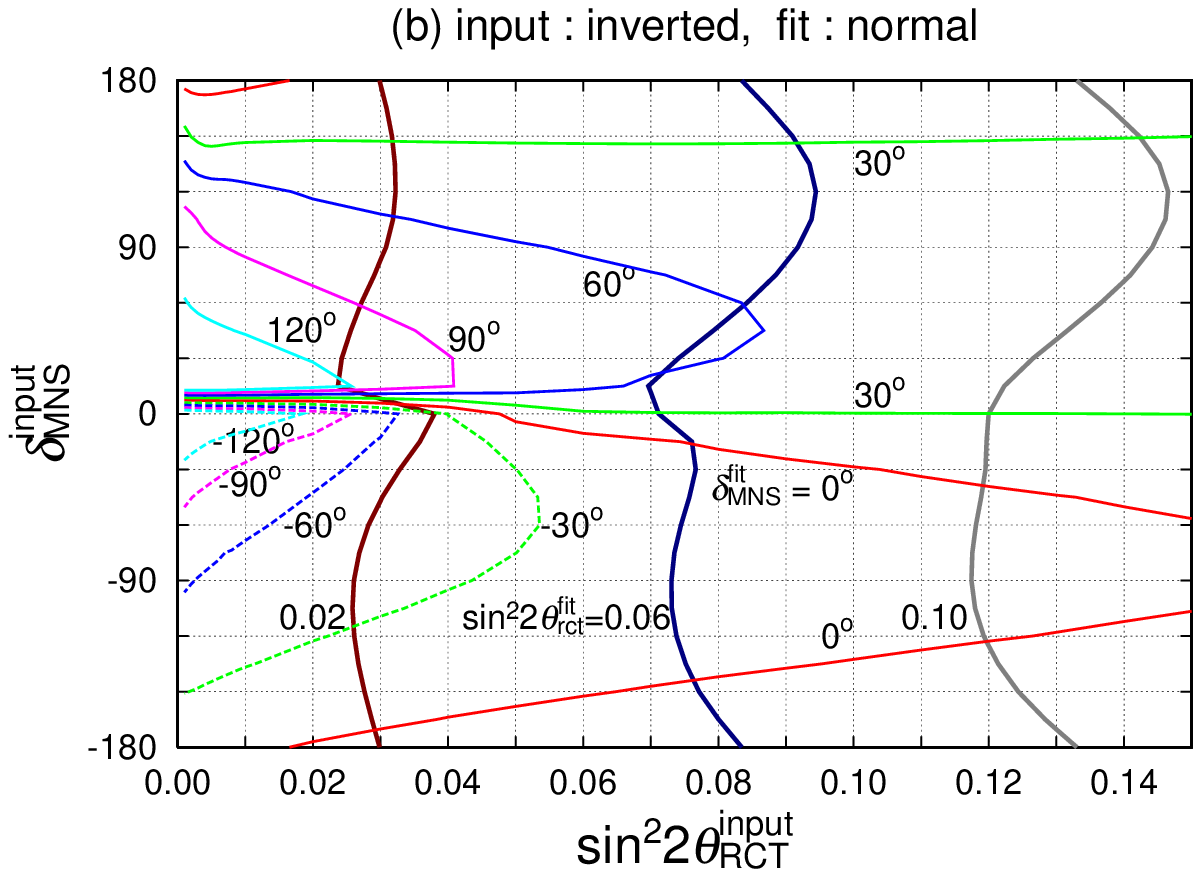}
\end{center}
\caption{
The values of the fit parameters, $\sir^{\rm fit}$ and $\dmns^{\rm fit}$,
at the minimum $\Delta \chi^2$ point of the analysis of \figref{fig:sensitivity} are
shown. The results for the normal hierarchy (a) and those for the inverted hierarchy (b)
are shown correspondingly to the fit of \figref{fig:sensitivity}~(a) and
\figref{fig:sensitivity}~(b), respectively.
The thick vertical lines are the $\sir^{\rm fit}$ contours at 0.02, 0.06, 0.10, and 0.14.
The thinner contours give the $\dmns^{\rm fit}$ values.
}
\label{fig:sensitivity-cp}
\end{figure}
This is clearly seen in \figref{fig:sensitivity-cp}, where
we show the values of the best fit parameters, $\sir^{\rm fit}$ and $\dmns^{\rm fit}$,
at the minimum $\Delta \chi^2$ point of the analysis of \figref{fig:sensitivity}.
The results for the normal hierarchy (a) and those for the inverted hierarchy (b),
are shown correspondingly to the fit of \figref{fig:sensitivity}(a)
and \figref{fig:sensitivity}(b),
respectively.
The thick vertical lines are the $\sir^{\rm fit}$ contours at 0.02, 0.06, 0.10
and 0.14.
The thinner contours give the $\dmns^{\rm fit}$ values.
We find that the value of $\dmns^{\rm fit}$ around $0^{\circ}$ is almost always favored as expected.
\par
Here let us try to explain more detailed features of \figref{fig:sensitivity} and \figref{fig:sensitivity-cp}
by separating the parameter space of $\sir^{\rm input}$ and $\dmns^{\rm input}$ into 4 regions.
\begin{enumerate}
\item small $\sir^{\rm input}$ region ( $\sir^{\rm input} < 0.04$) at
      any $\dmns^{\rm input}$:
In this region the phase difference \eqref{eq:Dphase} is mainly
      controlled by the $\cos \dmns$ terms,
because of the $1/\sqrt{\sir}$ enhancement over the matter effect term.
It is hence relatively easy to make the difference small by adjusting
     $\cos \dmns^{\rm fit} + \cos \dmns^{\rm input} \sim 0$.
The hierarchy is determined essentially by the difference of the
      $\nu_{\mu} \to \nu_e$ oscillation amplitude only.
\item $\sir^{\rm input} \simgt  0.04$ at $\dmns^{\rm input} \sim 180^{\circ}$:
Although the effect of $\cos \dmns^{\rm input} \sim -1$
is canceled by choosing $\cos \dmns^{\rm fit} \sim +1$,
the difference from the matter effect term in \eqref{eq:Dphase} cannot be canceled.
Therefore in this region the hierarchy is determined by the differences of both the amplitude and the oscillation phase.
\item $\sir^{\rm input} \simgt  0.04$ at $\dmns^{\rm input} \sim \pm 90^{\circ}$:
In \eqref{eq:Dphase}, the difference is controlled by the matter effect term and the $\cos \dmns^{\rm fit}$ term
because $\cos \dmns^{\rm input} \sim 0$.
In this region, we can make the phase-shift difference small by choosing $\cos \dmns^{\rm fit} > 0$.
\item $\sir^{\rm input} \sim 0.04$ to $0.08$ at $\dmns^{\rm input} \sim 0^{\circ}$:
In this region the phase-shift difference \eqref{eq:Dphase} at $|\Delta_{13}| \sim \pi$ can be
made small at $\cos \dmns^{\rm fit} \sim 1$,
but the difference at $|\Delta_{13}| \sim 2\pi$ becomes large.
Because the flux of \deg{0.5} off-axis beam is strong at lower energies where $\pi < |\Delta_{13}| < 2\pi$,
the growth of the phase-shift difference \eqref{eq:Dphase} at larger $|\Delta_{13}|$ cannot be compensated.
This explains why the minimum $\Delta \chi^2$ value
in this region is larger than the one for the case 3.
\end{enumerate} 
The systematics of the oscillation phase is rather complicated,
but its effect turns out to be significant in determining the neutrino mass hierarchy.
\par
Before closing the section,
let us briefly study the value of $\sir^{\rm fit}$ in \figref{fig:sensitivity-cp}.
In \figref{fig:sensitivity-cp}(a), $\sir^{\rm fit}$ is larger than $\sir^{\rm input}$,
whereas in \figref{fig:sensitivity-cp}(b), $\sir^{\rm fit}$ is smaller than $\sir^{\rm input}$.
This is because the same oscillation amplitude can be obtained by choosing $\sir^{\rm fit} > \sir^{\rm input}$
when the hierarchy is normal but it is assumed to be inverted in the fit,
and {\it vice versa} for the opposite case.
The fitted value of $\sir^{\rm input}$ cannot deviate too much from the input value of $\sir^{\rm input}$,
however, because of the constraint from the proposed reactor experiments,
according to the last term in \eqref{eq:chisq-para},
and also because of the SK measurement of the $\nu_{\mu} \to \nu_e$ probability,
which is much less sensitive to the mass hierarchy difference.
\section{Measurement of the CP phase}
\begin{figure}
\includegraphics[angle = 0 , width = 8cm] {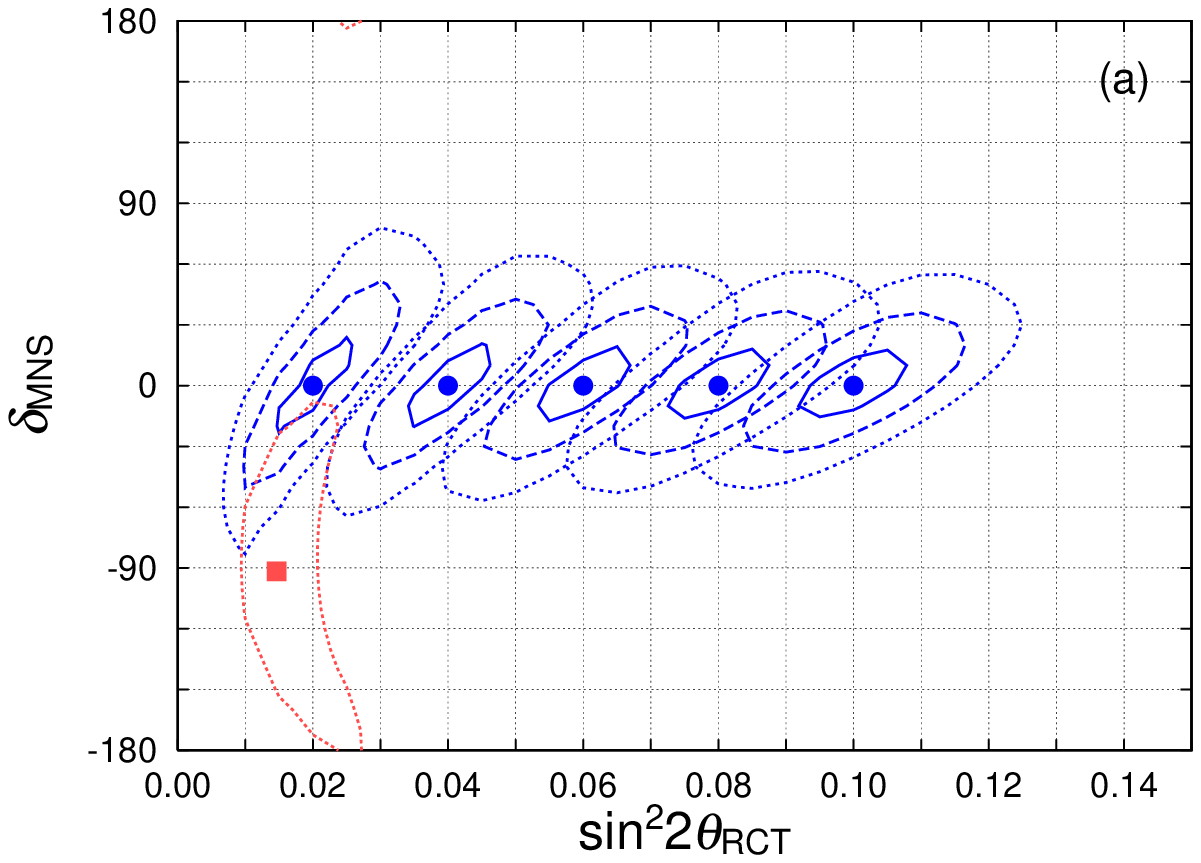}
\includegraphics[angle = 0 , width = 8cm] {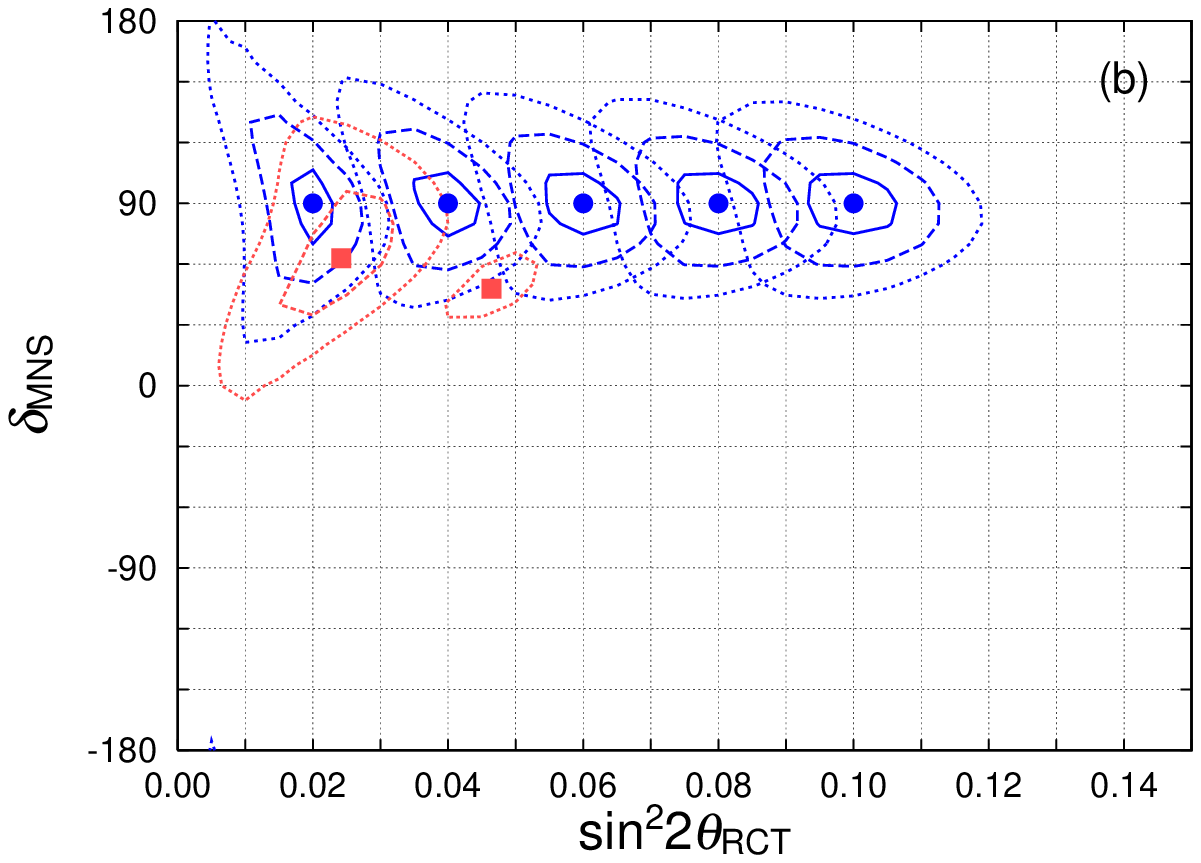}
\includegraphics[angle = 0 , width = 8cm] {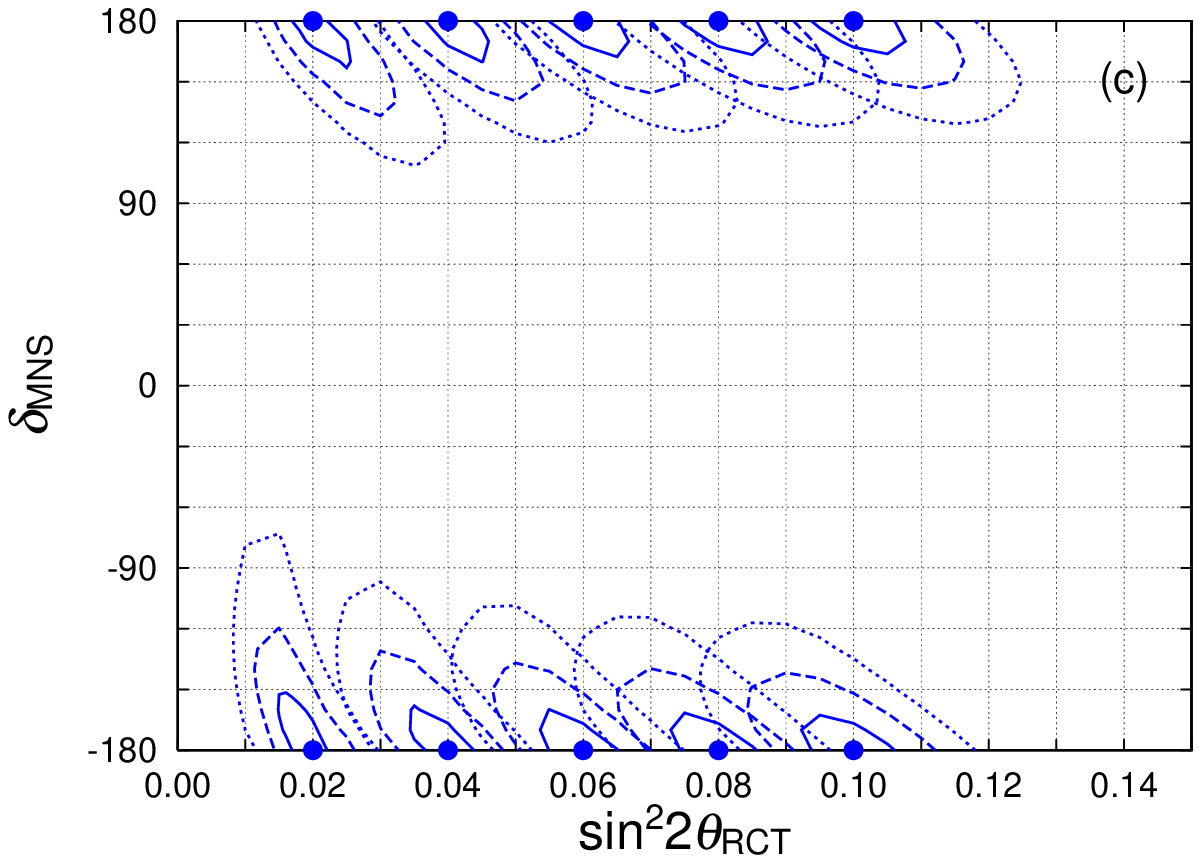}
\includegraphics[angle = 0 , width = 8cm] {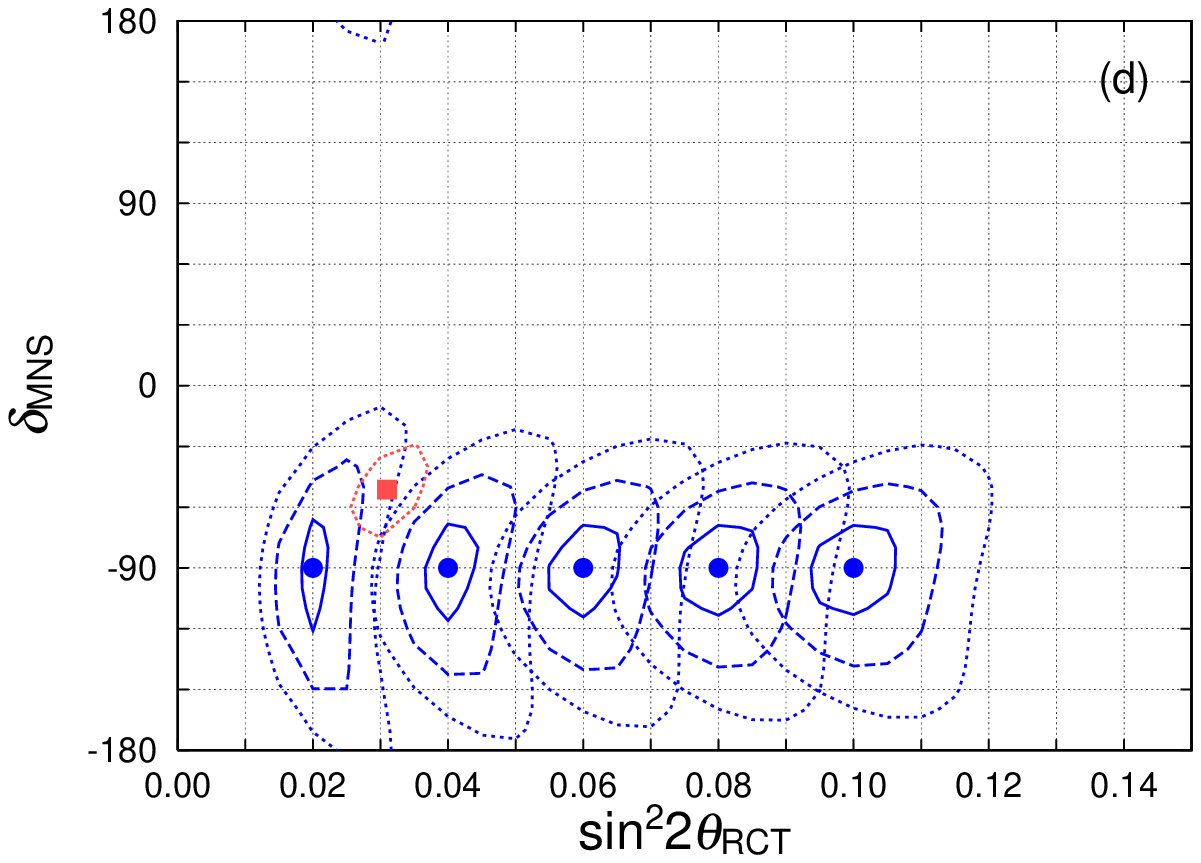}
\caption{
Capability of the T2KK two detector experiment for measuring \lsir and \ldmns.
Allowed regions in the plane of $\sir$ and $\dmns$
are shown for a combination of \deg{3.0}OAB at SK and \deg{0.5}at $L= 1000 \km$ with
a 100 kt water \cerenkov detector after 5 years of running ($5 \times 10^{21}$ POT).
The input values of \lsir are 0.02, 0.04, 0.06, 0.08 and 0.10 for
\ldmns = \deg{0}(a), \deg{90}(b), \deg{180}(c), and  \deg{-90}(d).
The normal hierarchy is assumed at  $m^2_3 - m^2_1 = 2.5 \times 10^{-3} {\rm eV^2}$,
and the 
other parameters are the same as those in Fig.~\ref{fig:sensitivity}.
The input points are shown as solid blobs, where $\Delta \chi^2 = 0$ by definition.
The 1-, 2-, and 3-$\sigma$ contours are then shown by solid, dashed, and dotted lines,
respectively.
For the input values of ($\sir^{\rm input}$, $\dmns^{\rm input}$)
= (0.02, \deg{0}) (a), (0.02, \deg{90}) and (0.04, \deg{90}) (b) and (0.02, \deg{-90}) (d),
there appear additional allowed regions when the mass hierarchy is chosen
with the wrong sign in the fit, where the local minimal $\Delta \chi^2$ point is depicted by
a solid square.
}
\label{fig:cp-normal}
\end{figure}
In this section we investigate the measurement of \lsir and \ldmns for
our preferred combination of the \deg{3.0} OAB at SK and \deg{0.5} OAB at $L = 1000$ km.
This combination of the T2KK experiment allows us to measure the $\nu_{\mu} \to \nu_e$ oscillation
around the oscillation maximum at two base-line lengths,
which can be parametrized as in \eqref{eq:p-numu-nue},
in terms of the amplitude shift
\eqref{eq:A-e+} and the phase shift \eqref{eq:B-e+}.
Once the neutrino mass hierarchy is determined as explained in the previous section,
the terms proportional to $\left| \Delta_{13} \right|/ \pi$
in the amplitude shift \eqref{eq:A-e+} measure $\sin \dmns$,
and those in the phase shift \eqref{eq:B-e+} measure $\cos \dmns$.
In the T2KK two detector system, both \lsir and $\sin \dmns$ can be determined
uniquely because the amplitude shift \eqref{eq:A-e+} has significantly different matter effect
contributions between SK and the far detector.
The phase shift measurement of the term \eqref{eq:B-e+} constrains $\cos \dmns$
independent of $\sin \dmns$.
\par
\begin{figure}[t]
\includegraphics[angle = 0 , width = 8cm] {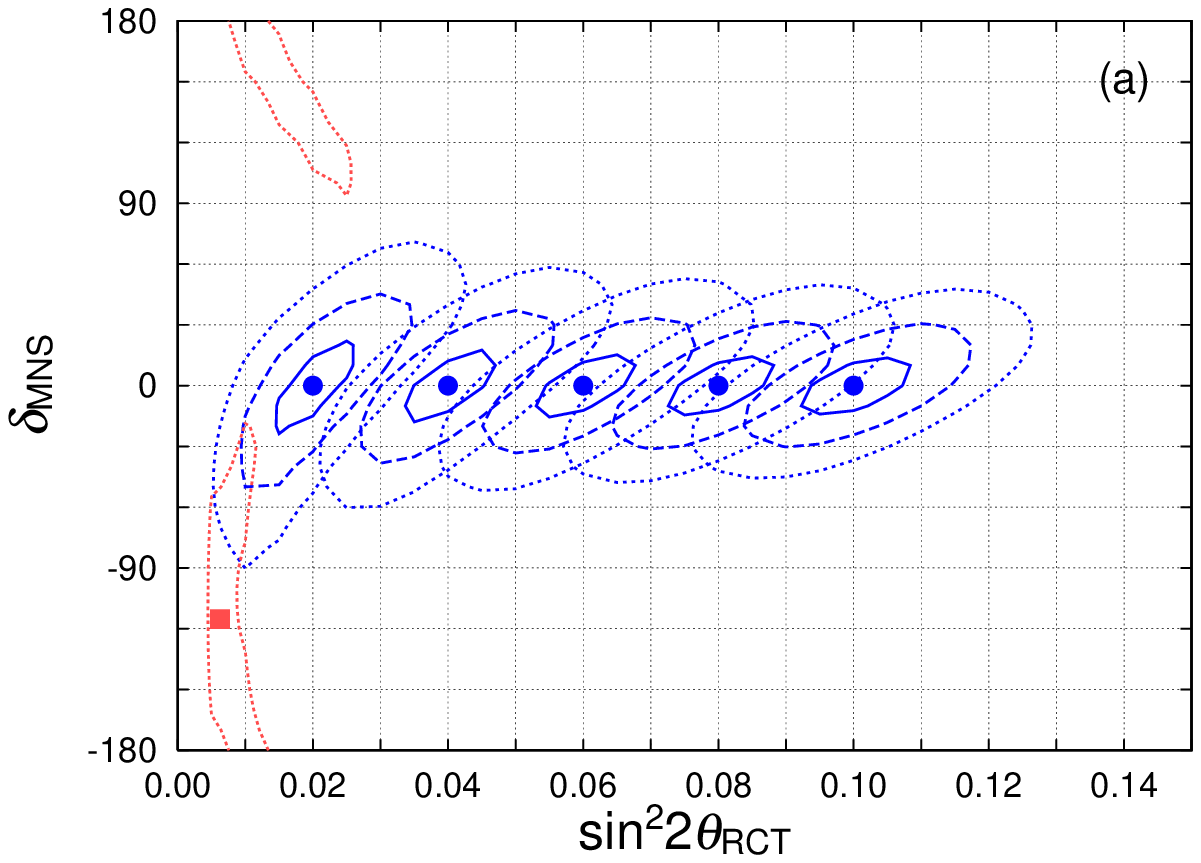}
\includegraphics[angle = 0 , width = 8cm] {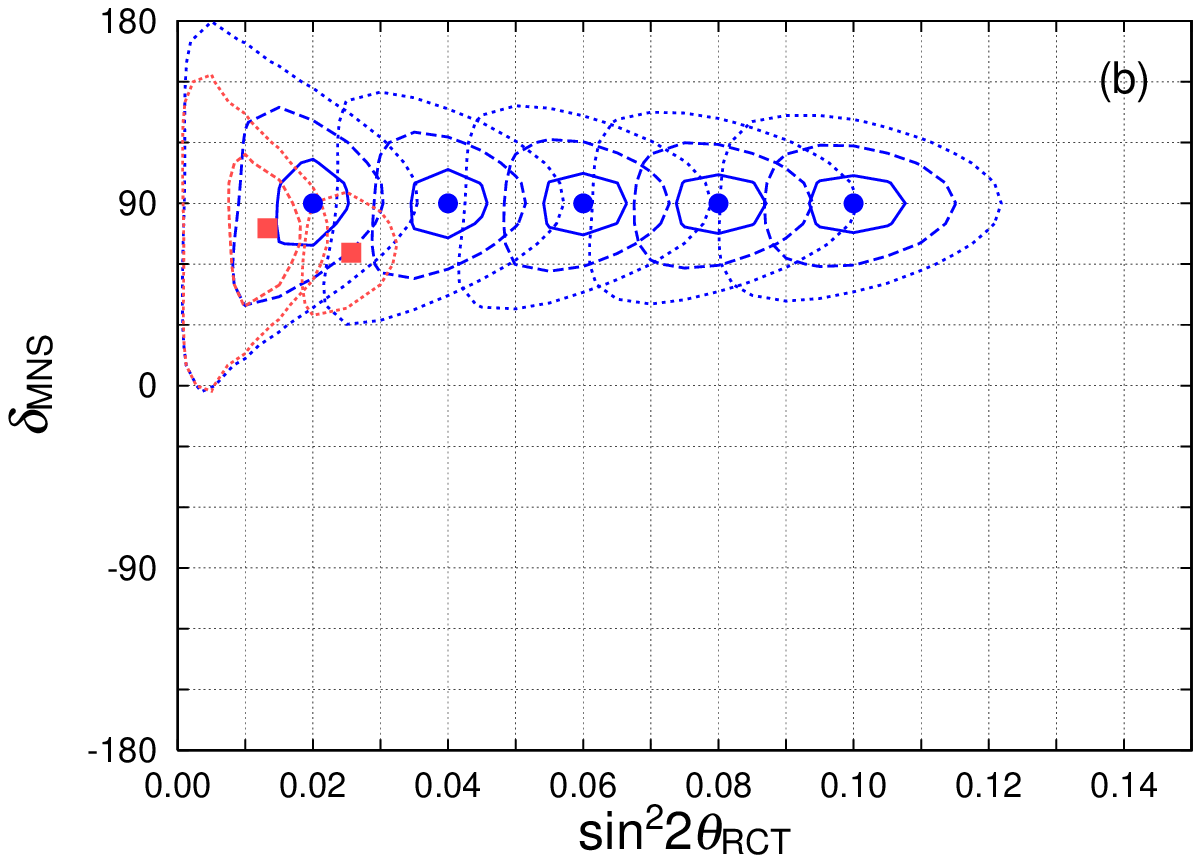}
\includegraphics[angle = 0 , width = 8cm] {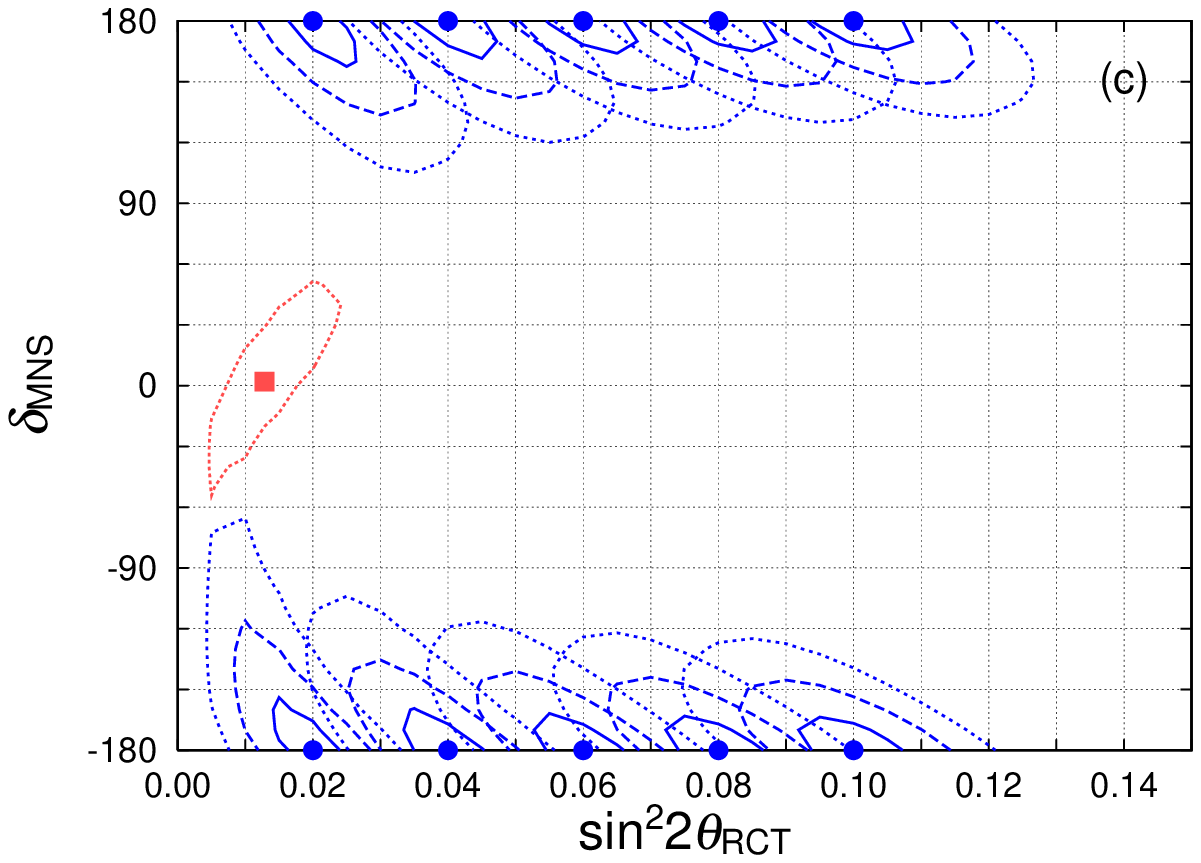}
\includegraphics[angle = 0 , width = 8cm] {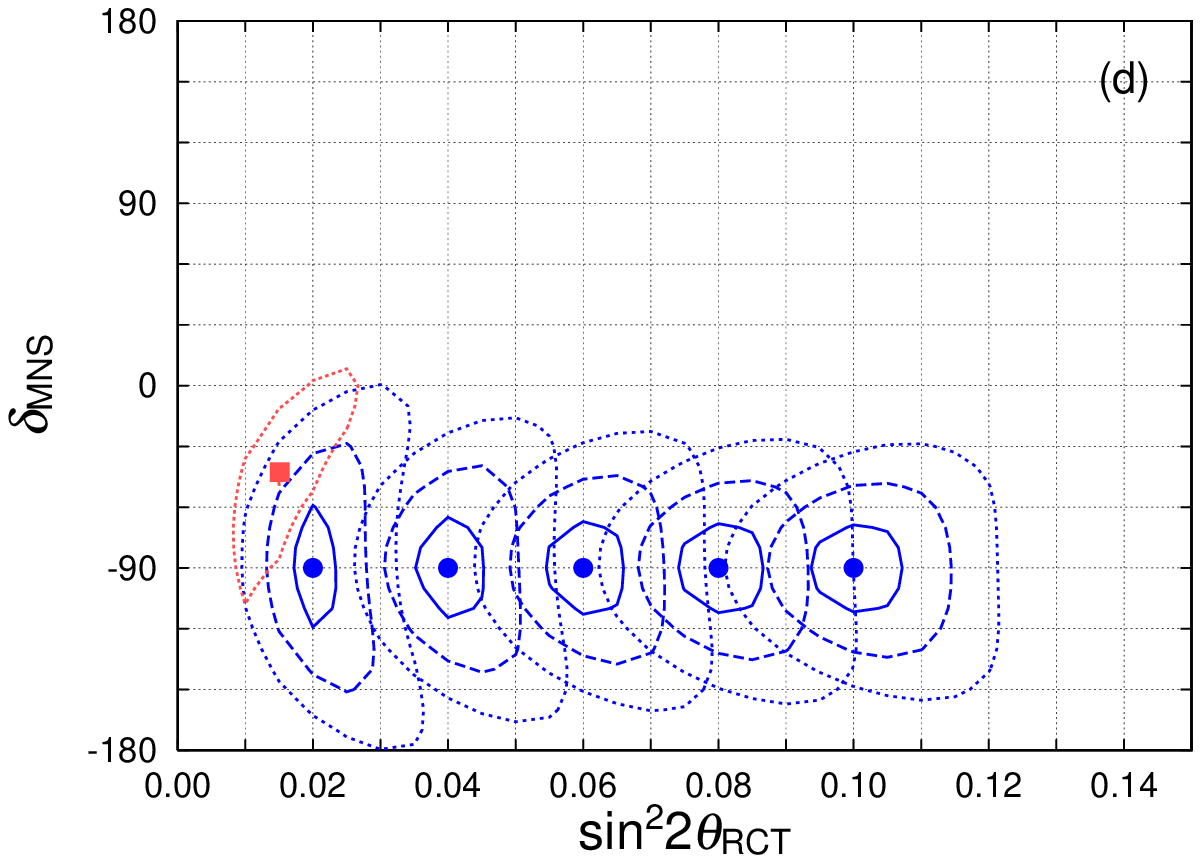}
\caption{
The same as \figref{fig:cp-normal}, but when the events are calculated for the inverted hierarchy,
{\it i.e.,} $m^2_3 - m^2_1 = - 2.5 \times 10^{-3} {\rm eV^2}$.
Just like in \figref{fig:cp-normal}, additional allowed regions,
when the wrong sign of the $m^2_3 - m^2_1$ is chosen in the fit,
appear for all the $\dmns^{\rm input}$ cases at $\sir^{\rm input} = 0.02$,
and for $\dmns^{\rm input} = 90^{\circ}$ (b) at $\sir^{\rm input} = 0.04$.
}
\label{fig:cp-inverted}
\end{figure}
We show in \figref{fig:cp-normal} and \figref{fig:cp-inverted} regions
allowed by this experiment
in the plane of \lsir and $\dmns$.
The mean values of the input data are calculated for the parameters 
of \eqref{eq:input}.
In each figure,
input points ($\sir^{\rm input}, \dmns^{\rm input} $) are
shown by solid-circles for $\sir^{\rm input}$ between 0.02 and 0.1,
with an interval of 0.02,
and for four values of $\dmns^{\rm input}$;
\deg{0}(a), \deg{90}(b), \deg{180}(c), and $-90^{\circ}$(d).
The regions where the minimum $\Delta \chi^2$ value is less than 1, 4, 9
are depicted by solid, dashed, and dotted boundaries, respectively.
\figref{fig:cp-normal} is for the normal hierarchy,
and \figref{fig:cp-inverted} is for the inverted hierarchy.
From these figures,
we find that $\dmns$ can be constrained to $\pm 30^{\circ}$ at
1-$\sigma$ level,
when $\sir^{\rm input} \simgt 0.02$ as long as the neutrino mass
hierarchy is determined.
If we remove the last term in eq.(\ref{eq:chisq-para}),
the error of \ldmns changes little but that of \lsir
grows significantly; see Fig. 5 of Ref.~\cite{t2kr-l}
\par
As shown in \figref{fig:sensitivity}(a) and (b),
the mass hierarchy cannot be determined at 3-$\sigma$ level
($\Delta \chi^2 >$ 9) when $\sir^{\rm input}$ is too small.
In case of the input parameters of \figref{fig:cp-normal} for the
normal hierarchy,
this is the case for $\sir^{\rm input} = 0.02$ at $\dmns^{\rm input} =
0^{\circ}$ (a),
$\sir^{\rm input} = 0.02$ and 0.04 at $\dmns^{\rm input} = 90^{\circ}$ (b),
and $\sir^{\rm input} = 0.02$ at $\dmns^{\rm input} = - 90^{\circ}$(d).
For those input points,
there appear an additional allowed region whose center (local minimum
of $\Delta \chi^2$) is shown by a solid square.
No extra allowed region appears for $\dmns = 180^{\circ}$ in
\figref{fig:cp-normal}(c),
in accordance with the result of \figref{fig:sensitivity}(a).
In case of \figref{fig:cp-inverted} for the inverted hierarchy,
the local minimum appears for
$\sir^{\rm input} = 0.02$ at $\dmns^{\rm input} = 0^{\circ}$ (a),
$\sir^{\rm input} = 0.02$ and 0.04 at $\dmns^{\rm input} = 90^{\circ}$ (b),
$\sir^{\rm input} = 0.02$ at $\dmns^{\rm input} = 180^{\circ}$ (c),
and
$\sir^{\rm input} = 0.02$ at $\dmns^{\rm input} = - 90^{\circ}$(d).
\par
It is remarkable that the error of $\dmns$ is almost independent of $\sir^{\rm input}$
value between 0.02 and 0.1,
for all the four input values of $\dmns^{\rm input}$, \deg{0}, $\pm$\deg{90}, and \deg{180}.
This is remarkable because the event number $N_e$ is proportional to \lsir according to \eqref{eq:p-numu-nue},
and hence the statistical error of the measurement of the amplitude and the phase should be proportional to $1/\sqrt{N_e}$,
or $1/\sqrt{\sin^2 \theta_{\rm \footnotesize RCT}}$.
This increase in the error for small $\sir^{\rm input}$ values
is canceled by the increased sensitivities of both
the amplitude and the phase shift to $\sin \dmns$ and $\cos \dmns$,
respectively,
which are both proportional to $1/\sqrt{\sin^2 \theta_{\rm \footnotesize RCT} }$.
The two effects cancel rather accurately, and we find that the error of $\dmns$
is almost independent of the input
values of $\sir$ and $\dmns$.
\section{Conclusion and discussions}
In this paper we study physics potential of the T2KK proposal
\cite{hagiwara-see-saw, t2kk, t2kr-l},
where a far detector along the T2K neutrino beam line is placed in Korea.
We find that the off-axis neutrino beam from J-PARC at Tokai village
for the T2K
project has significant intensity at a few GeV range
when the far detector is placed in the east coast of Korea where
the beam at less than \deg{1.0} off-axis can be observed at a base-line
length of $L \sim 1000$ km.
The resulting two detector system can observe the $\nu_{\mu} \to \nu_e$
oscillation probability near the oscillation maximum at two different energies,
if \lsir is not too small.
We examine,
in particular, the capability of determining the mass hierarchy
pattern and the CP phase of the lepton-flavor-mixing matrix
when a 100 kt water \cerenkov detector is placed at various locations in Korea
for the off-axis beam (OAB) of \deg{2.5} and \deg{3.0} at the
Super-Kamiokande site.
The best results are found for a combination of \deg{3.0} OAB at SK ($L
= 295 \km$)
and \deg{0.5} OAB at $L = 1000 \km$,
where the mass hierarchy pattern can be determined at 3-$\sigma$ level
for $\sir  \simgt 0.05$ $(0.06)$ when the hierarchy is normal (inverted), after 5 years of
running ($5 \times 10^{21}$ POT).
The sensitivity of the T2KK experiment on the neutrino mass hierarchy
depends not only on \lsir but also on \ldmns.
We explore the sensitivity in the whole space of \lsir and \ldmns,
and the results are shown in \figref{fig:sensitivity}(a) for the normal
hierarchy and in \figref{fig:sensitivity}(b) for the inverted hierarchy.
Significantly higher sensitivity is found for $\dmns \sim 180^{\circ}$
for both hierarchy cases.
\par
We also find that the leptonic CP phase, \ldmns, can be constrained uniquely,
without invoking anti-neutrino beams,
as long as the mass hierarchy pattern is determined: see \figref{fig:cp-normal}
for the normal hierarchy,
and \figref{fig:cp-inverted}
for the inverted hierarchy.
\par
Those results are obtained by assuming that the neutrino energy can be
reconstructed with a hundred MeV uncertainty for the charged current
quasi-elastic events, and the earth matter density along the baseline
can be determined with $3\%$ accuracy.
All our numerical results have been understood semi-quantitatively,
by using the approximative expression for the $\nu_{\mu} \to \nu_{\mu}$
and $\nu_{\mu} \to \nu_e$ oscillation
probabilities,
eqs.~(\ref{eq:p-numu-numu}),
(\ref{eq:AB-mu}),
(\ref{eq:p-numu-nue})
(\ref{eq:AB-e}),
and
(\ref{eq:factorize}),
where only the linear and quadratic terms
in the matter effect and those in \lsir are
kept in the oscillation probabilities.
\par
Our results are based upon a very simple treatment of the systematic 
errors where $3\%$ overall errors are assigned for all the 10 normalization 
factors of eq.~(\ref{chisq-sys}).  
We find that the significance of the mass hierarchy determination is not affected much
even if we enlarge all the systematic errors to $10\%$ except for the matter density uncertainties.
This means that the errors of the T2KK experiment proposal
in this paper are dominated by the statistical error.
Therefore, if we make the detectors at Kamioka and{\footnotesize /}or Korea larger,
or if the intensity of the J-PARC beam is stronger,
the significance of the
measurement will grow as
proportional to the square root of the product of the volume,
the intensity, and the exposure time.
It is not so clear,
however,
what is the best volume ratio between the detector in Kamioka and that in Korea.
\begin{figure}[t]
\begin{center}
\includegraphics[angle = 0 ,width=7.5cm]{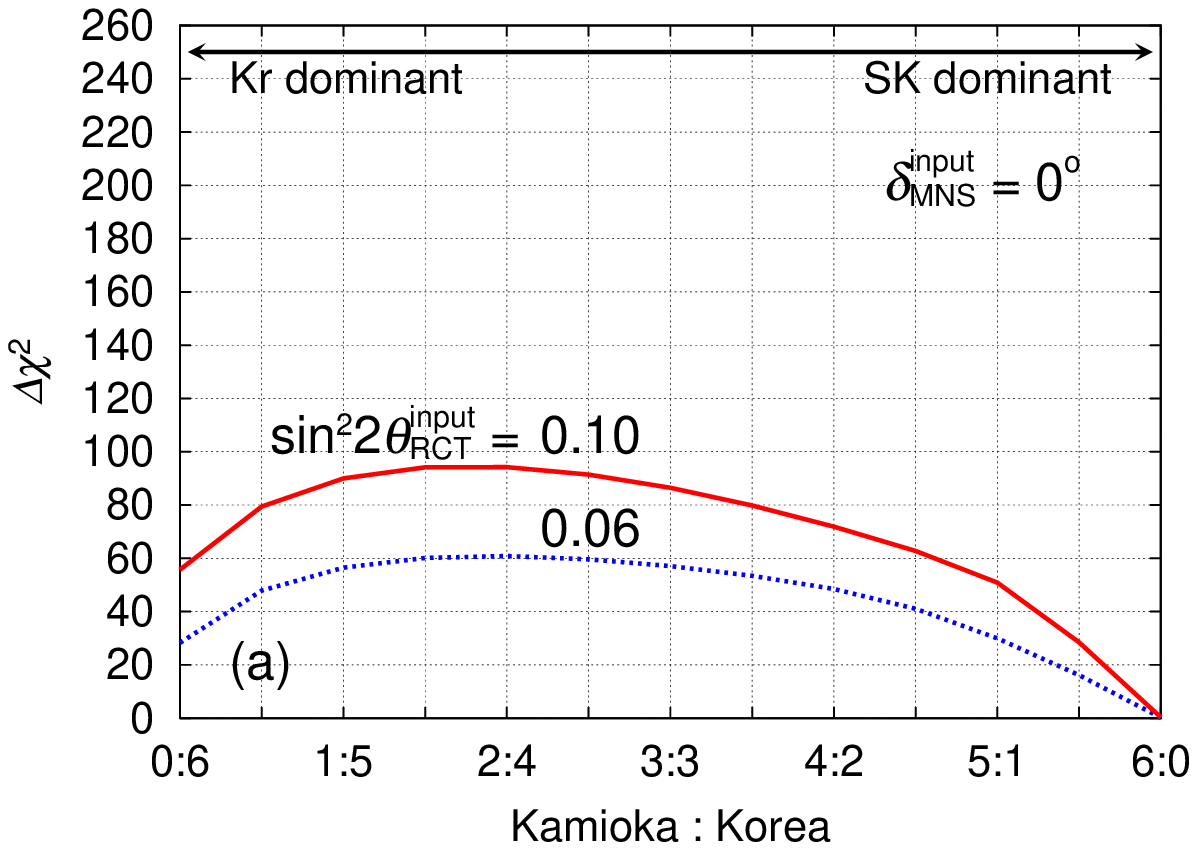}
\includegraphics[angle = 0 ,width=7.5cm]{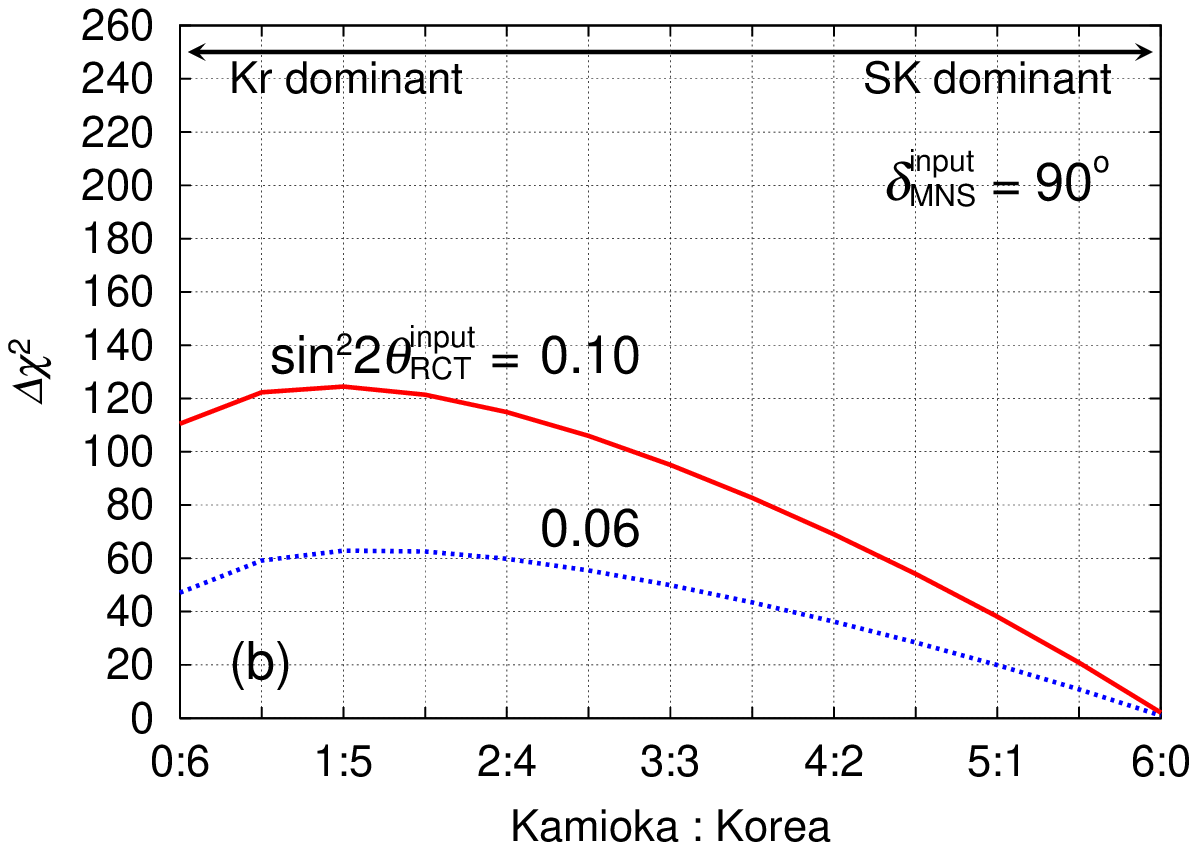}
\includegraphics[angle = 0 ,width=7.5cm]{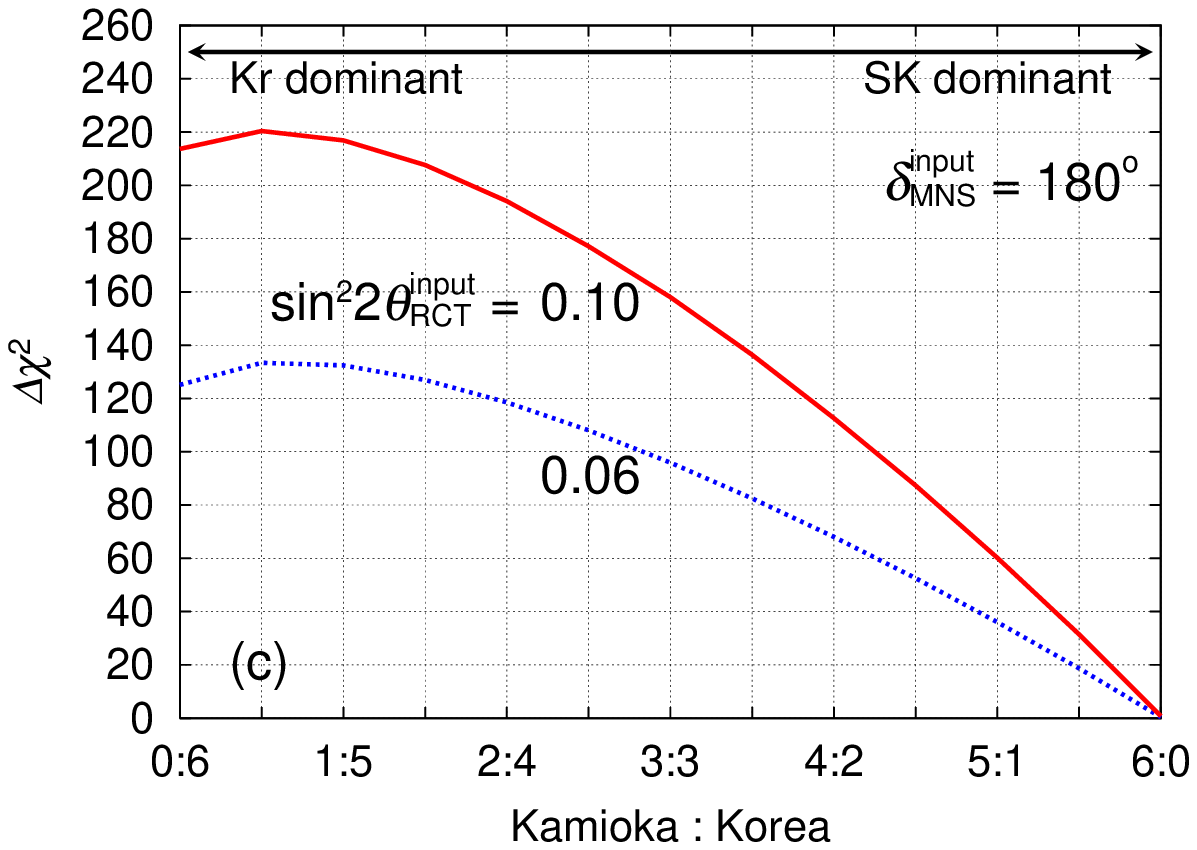}
\includegraphics[angle = 0 ,width=7.5cm]{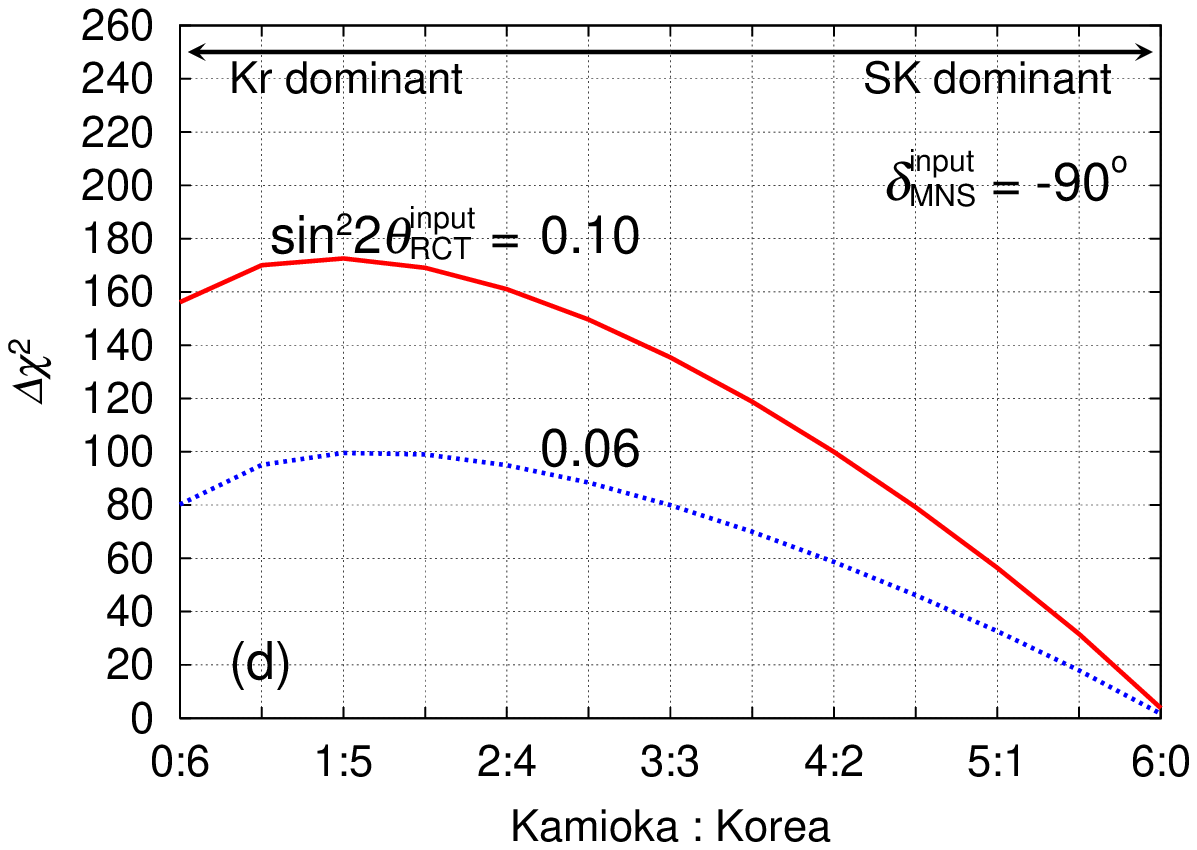}
\end{center}
\caption{
Optimal ratio of the fiducial volumes of two detectors, one at Kamioka ($L = 295$ km)
and the other at $L = 1000$ km in Korea, for determining the neutrino mass hierarchy.
The input data are calculated for the normal hierarchy at
$m^2_3 - m^2_1 = 2.5 \times 10^{-3} {\rm eV}^2$ and the minimum $\Delta\chi^2$
of the fit with the wrong hierarchy is shown for $\sir^{\rm input} = 0.10$ (solid lines)
and 0.06 (dotted lines) and for $\dmns^{\rm input} =$ \deg{0} (a), \deg{90} (b),
\deg{180} (c), and \deg{270} (d),
when the sum of the fiducial volumes is fixed at 600 kt.
The other parameters are same as those in \figref{fig:flux} (d),
and the results are calculated for $5 \times 10^{21}$POT.
}
\label{fig:volume}
\end{figure}
We show in \figref{fig:volume} the minimum
$\Delta\chi^2$ as functions of the volume ratio of the near (Kamioka) and far (Korea) detectors
while keeping the total volume at 600 kt for $5 \times 10^{21}$ POT.
We assume the normal hierarchy for the input and the inverted hierarchy
in the fit.
We examine 8 cases,
for $\sir^{\rm input} = 0.1$ (solid lines) and
$\sir^{\rm input} = 0.06$ (dotted lines),
and for $\dmns^{\rm input}$ = \deg{0} (a),
\deg{90} (b), \deg{180} (c), and \deg{-90}(d).
It is clearly seen from \figref{fig:volume}
that a 600 kt detector in Kamioka alone cannot resolve the mass
hierarchy at all,
because there is a little difference in $\nu_{\mu} \to \nu_e$
transition probability between the normal hierarchy
and the inverted hierarchy.
On the other hand,
in the case of only a Korean detector with 600 kt,
the minimum $\Delta \chi^2$ value is not much smaller than the best case.
This is because the constraint of \lsir from the future reactor
neutrino experiment replaces the role of the near detector which
measures the $\nu_{\mu} \to \nu_e$ transition
at low energies where the matter effect is small.
We find that the minimum $\Delta \chi^2$ value of 23.5
in \figref{fig:volume}(a)
at the volume ratio of 1:5 ($\approx 22.5 : 100$)
is about 4.5 times as large as the minimum $\Delta \chi^2$ value
in \figref{fig:place-n}(b), confirming the
dominance of the statistical error in our analysis.
If we request that the minimum $\Delta \chi^2$ should be at least $80\%$
of its optimal value,
then the near-to-far volume ratio
should be between $0.5:5.5$ and $2.5 : 3.5$.
More volumes should be given to the far detector than to the near detector.
The above results are obtained with a $100\%$ detection efficiency for the CCQE
events. If we set $80\%$ efficiency for SK and $60\%$ efficiency for a far detector
in Korea \cite{nakayama}, the value of $\Delta \chi^2$ in \figref{fig:place-n}
for the combination of $3.0^{\circ}$ for SK and $0.5^{\circ}$ for a far detector placed
at $L= 1000$ km drops from 22.6 to 14.5, roughly scaling with the effective number of events
at a far detector.
\par
We note here that our proposal is significantly different from that
of Ref.\cite{t2kk} in the following aspects.
\begin{enumerate}
\item We propose to use a combination of large off-axis angle at SK and
a small off-axis angle at the far detector, so that the first
oscillation maximum can be observed at both detectors.
\item As a consequence, the CP phase can be determined uniquely without
invoking experiments with an anti-neutrino beam.
\item We propose that the far detector should have larger fiducial volume
in order to maximize the physics outputs, see \figref{fig:volume}.
\end{enumerate}
The use of two different off-axis angles implies that our proposal
should suffer from larger systematic errors, especially for high
energy events at the far detector.
Among the potentially serious background which we could not estimate
in this paper are;
\begin{itemize}
  \item possible miss-identification of NC $\pi^0$ production as
	$\nu_e$ CCQE events,
  \item possible miss-identification of soft $\pi$ emission events as
	CCQE events.
\end{itemize}
Although the above uncertainties were found to be rather small at K2K
experiments \cite{NC},
we should expect them to be more serious at high energies.
Dedicated studies on the fake reconstructed neutrino energy of the NC $\pi^0$
events and those on the correlation between the true and the reconstructed
neutrino energy of the CC soft $\pi$ events are mandatory.
In addition,
careful studies including possible energy 
dependence of the flux and cross section uncertainties, and also the location 
dependence of the matter density, may be needed to justify the physics
case of the T2KK proposal.
\vspace{0.5cm}
\\
{\it Acknowledgments}\\
We thank our colleagues
Y.~Hayato,
A.K.~Ichikawa,
T.~Ishii,
I.~Kato,
T.~Kobayashi
and
T.~Nakaya,
from whom we learn about the K2K and T2K experiments.
We are also grateful to 
Mayumi Aoki,
Paul H.~Frampton,
Chung-Wook Kim,
Soo-Bong Kim,
and
Yeongduk Kim,
for useful discussions and comments.
The work is supported in part
by the Core University Program of JSPS.
The numerical calculations were partly carried out on
Altix3700 BX2 at YITP in Kyoto University.

\end{document}